\documentclass[prb,showpacs,preprintnumbers,twocolumn,amsmath,amssymb,superscriptaddress]{revtex4}
\usepackage{graphicx}
\usepackage{dcolumn}
\usepackage{bm}
\usepackage{psfrag}
\usepackage{color}

\def\bR{{\bf R}}

\begin{document}
\title{Amplitude, density and current correlations of strongly disordered 
superconductors} 
\author{G.~Seibold} 
\affiliation{Institut F\"ur Physik, BTU Cottbus-Senftenberg, PBox 101344, 03013 Cottbus,
Germany}
\author{L.~Benfatto} 
\affiliation{ISC-CNR and Department of Physics, University of Rome ``La
  Sapienza'',\\ Piazzale Aldo Moro 5, 00185, Rome, Italy}
\author{C.~Castellani}
\affiliation{ISC-CNR and Department of Physics, University of Rome ``La
  Sapienza'',\\ Piazzale Aldo Moro 5, 00185, Rome, Italy}
\author{J.~Lorenzana}
\affiliation{ISC-CNR and Department of Physics, University of Rome ``La
  Sapienza'',\\ Piazzale Aldo Moro 5, 00185, Rome, Italy}

\date{\today}

\begin{abstract}
We investigate the disorder dependence of the
static density, amplitude and current correlations 
within the attractive Hubbard model supplemented with on-site
disorder. It is found that strong disorder favors a decoupling of 
density and amplitude correlations
due to the formation of superconducting islands.
This emergent granularity also induces an enhancement of the 
density correlations on the SC islands whereas amplitude fluctuations are
most pronounced in the 'insulating' regions. While density and
amplitude correlations are short-ranged at strong disorder we show
that current correlations have a long-range tail due to the 
formation of percolative current paths in agreement with the constant
behavior expected from the analysis of one-dimensional models.
\end{abstract}


\maketitle
\section{Introduction}
More than $50$ years ago Anderson has discussed the behavior of
a superconductor in the presence of strong disorder.~\cite{and59}
According to his analysis (and under the restriction to elastic scattering
from non-magnetic impurities) the BCS wave-function, build from Bloch-type 
wave functions with opposite momenta, can be generalized to pairs made
from the exact single-particle wave-functions of the disordered system
plus their time-reversed partner.
As a result one would expect a gradual dependence of the superconducting
transition temperature on the presence of non-magnetic impurities
caused mainly by a modification of single-electron properties as density
of states etc. While this picture is certainly correct for weak
disorder, experiments on thin films of strongly disordered 
superconductors \cite{hav89,heb90,shahar92,adams04,steiner05,stew07,sac08,sac10,mondal11,chand12,kaml13,noat13} have 
revealed
a much more interesting behavior than suggested by Ref.~\onlinecite{and59}.
In particular, the observation of a superconductor-insulator transition 
(SIT) with increasing disorder provides evidence for an interesting interplay 
between localization of Cooper pairs and long-range superconducting 
(SC) order.~\cite{fm10} 
Moreover, the observation of a pseudogap in strongly disordered SC
films \cite{sac08,sac10,mondal11,chand12,kaml13,noat13} bears some resemblance to 
similar
experimental findings in high-temperature superconductors 
\cite{timusk,fischer07,rullier11} that may suggest a 
common mechanism in some regions 
of the phase diagram.

Theoretical investigations of disordered superconductors are either
based on bosonic or fermionic approaches. In case of s-wave superconductivity
the latter typically start from attractive Hubbard models where disorder 
is usually implemented via a shift of onsite energy levels.~\cite{triv96,ghosal98,scal99,ghosal01,dubi2007,dey2008,boua11,sei12,ghosh13} These hamiltonians then 
are either treated within 
a standard Bogoljubov-de Gennes approximation 
\cite{ghosal98,ghosal01,dubi2007,dey2008,sei12,ghosh13}
or with more  sophisticated approaches 
like Monte-Carlo methods.~\cite{triv96,scal99,boua11}
Bosonic models are then obtained from a large-U expansion, as e.g. 
the pseudospin $XY$ model in a transverse field\cite{ma_prb85},  where the
hopping of Cooper pairs (corresponding to pseudospins aligned in the $XY$ plane) competes with localization
due to random fields (corresponding to pseudospins aligned in the perpendicular direction). Further simplifications, as e.g. an Ising model in a 
random transverse field, are also introduced since they allow for 
analytical treatments.~\cite{mez10}

In recent years both approaches have lead to a coherent picture of the
SIT: With increasing disorder the system starts to break up into
``puddles'' with finite SC order parameter $|\Delta| > 0$ and 
intermediate regions with $|\Delta| \approx 0$ although the spectral gap
remains finite. The order parameter
distribution shows a universal scaling behavior, in agreement with
experiment, where the relevant scaling variable is the logarithm of the
order parameter distribution normalized to its variance.~\cite{lem13,mayoh15}
The phases of different puddles are weakly coupled, so that the
system bears some resemblance with a granular superconductor.
Upon applying a vector potential the system accommodates the
phase twist in the regions with $|\Delta| \approx 0$ so that the 
associated energy, and thus the superfluid stiffness, are strongly
reduced. Moreover, calculations within the BdG approach of the
attractive Hubbard model have shown that the induced current
flows along a quasi one-dimensional percolative path or
``superconducting backbone''  which 
connects the puddles.~\cite{sei12} This result has its counterpart in
the analysis of the bosonic approach which has revealed a regime 
of broken-replica symmetry
where the partition function is determined by a small number of 
paths.~\cite{mez10} 
For both, fermionic and bosonic models, there
exists a critical value for the disorder strength above which
the system becomes insulating. The SIT is characterized by a vanishing
of the superfluid stiffness, however, the single-particle gap persists
across the transition\cite{boua11}. 

A still open issue is the nature of the spatial correlations in such granular SC state arising near the SIT. In the classical Ginzburg-Landau-Abrikosov-Gorkov theory \cite{glag} there is a single scale $\xi_0\sim v_F/\Delta$,  whose reduction by disorder is mainly governed by the mean-free path $\ell$ via 
$\xi\sim \sqrt{\xi_0\ell}$. On the other hand, in the vicinity of the Anderson
localization transition the coherence length is also controlled by the 
localization length.~\cite{kapi85,kotliar86} 
Concerning the disordered attractive Hubbard model with a
fragmented SC ground state as  mentioned above,
there is only limited knowledge about amplitude, 
density and current correlations.
Previous Quantum Monte-Carlo studies \cite{scal99} yield only 
limited information on the spatial 
dependence of the correlations due to the small ($8\times 8$) lattice
sizes.  
On the other hand investigations of response functions
on larger clusters within the
BdG approach where so far restricted to mean-field studies.   
In the present paper we evaluate the density, amplitude and current 
correlations by including fluctuations on top of
the BdG solution thus generalizing the approach of 
Refs.~\onlinecite{strinati96,lara}
to the case with disorder. In particular we are interested in the
question of how the physics is governed by different length scales in
different channels and how the formation of SC islands for strong disorder
reflects in the corresponding correlation lengths.   
  
The paper is organized as follows:
The model is introduced in Sec.~\ref{sec:form} where we also outline
the computation of correlation functions on the basis of the BdG ground state.
Results are presented in Sec.~\ref{sec:res} for amplitude,
density and current correlations. We finally conclude our discussion in Sec.~\ref{sec:conc}.

\section{Formalism}\label{sec:form}
\subsection{BdG equations}
Our starting point is the attractive Hubbard model with local
disorder                                                 
\begin{equation}                                         
H=\sum_{ij\sigma}t_{ij}c^\dagger_{i\sigma}c_{j\sigma} -|U|\sum_{i}n_{i\uparrow}n_
{i\downarrow} +\sum_{i\sigma}V_i n_{i\sigma}                                      
\end{equation}                                                                    
which we solve in mean-field using the BdG transformation
\begin{displaymath}
c_{i\sigma}=\sum_k\left[u_i(k)\gamma_{k,\sigma}-\sigma v_i^*(k)\gamma_{k,-\sigma
}^\dagger\right]
\end{displaymath}
\begin{align}
\omega_k u_n(k)&=\sum_{j}t_{nj} u_j(k) + [V_n-\frac{|U|}{2}\langle n_n\rangle -
\mu] u_n(k) \nonumber \\ &+\Delta_n v_n(k)                              
 \label{eq1}\\                                                                  
\omega_k v_n(k)&=-\sum_{j}t^*_{nj} v_j(k) -
[V_n-\frac{|U|}{2}\langle n_n\rangle -\mu] u_n(k)\nonumber \\
&+\Delta^*_n u_n(k)\label{eq2}\,.
\end{align}

For simplicity only nearest-neighbor hopping
$t_{ij}=-t$  is considered
in this work. The disorder variables $V_i$ are taken from a flat, normalized
distribution ranging from $-V_0$ to $+V_0$.
 
In the following $u_i(k)$ and $v_i(k)$ are taken to be real.
Starting from an initial distribution of the gap $\Delta_i$ and
density $\langle n_i\rangle$ values we
diagonalize the system of equations (\ref{eq1},\ref{eq2}),
compute the new values ($T=0$)
\begin{eqnarray}
\label{op}
 \Delta_i&=&|U|\sum_n u_i(n)v^*_i(n) \\
\langle n_i\rangle &=& 2\sum_n|v_i(n)|^2
\end{eqnarray}
and iterate the obtained values, say $K$, 
(including also the chemical potential)
up to a given accuracy $\delta K/K \le \epsilon$, typically $\epsilon=10^{-6}$.
For the disordered systems studied in Sec.~\ref{sec:res} clusters 
with up to $24\times 24$ sites have been diagonalized. We mostly show
results with filling $n=0.875$, but in some cases we also discuss the outcomes
for smaller filling in order to avoid the proximity to half-filling, 
where specific effects can arise due to the tendency of the system to 
form a charge-density-wave (CDW) state as well.

\subsection{Amplitude and Charge Correlations}

We denote correlation functions by
\begin{equation}
\chi^{O,R}_{nm}(\omega)= i\int\!dt e^{i\omega t}
\langle {\cal T} \hat{O}_n(t) \hat{R}_m(0)\rangle
\end{equation}
where in the following $\hat{O}$,$\hat{R}$ correspond
to either amplitude 
$\delta A_i$ or density $\delta \rho_i$ fluctuations
\begin{eqnarray*}
\delta A_i &\equiv & (\delta\eta_i+\delta\eta^\dagger_i)/\sqrt{2} \\
\delta \rho_i &\equiv& \sum_{\sigma}\left(c^\dagger_{i\sigma}c_{i\sigma} - \langle c^\dagger_{i\sigma}c_{i\sigma} \rangle\right)\,,
\end{eqnarray*}
and we have defined the pair fluctuation operators 
\begin{eqnarray*}
\delta\eta_i^\dagger &\equiv& c^\dagger_{i\uparrow}c^\dagger_{i\downarrow} - 
\langle c^\dagger_{i\uparrow}c^\dagger_{i\downarrow} \rangle \\
\delta\eta_i &\equiv& c_{i\downarrow}c_{i\uparrow} - \langle c_{i\downarrow}c_{i\uparrow} \rangle \,.
\end{eqnarray*} 

It is then convenient to define $2\times 2$ matrices for the 
bare mean-field susceptibility

\begin{equation}
\underline{\chi^0}_{ij}=\left(
\begin{array}{cc}
\chi^{AA}_{ij} & \chi^{A\rho}_{ij} \\
\chi^{\rho A}_{ij} & \chi^{\rho\rho}_{ij}
\end{array}\right) .
\end{equation}

and the interaction

\begin{equation}
\underline{V}=\left(
\begin{array}{ccc}
-|U| & 0 \\
0 & -|U|/2
\end{array}\right)
\end{equation}

which can be combined into ``large'' matrices according to

\begin{equation}
\underline{\underline{\chi^0_{ij}}} = \left(
\begin{array}{cccc}
\underline{\chi^0_{11}} & \underline{\chi^0_{12}} & \cdots & \underline{\chi^0_{1N}} \\ 
\underline{\chi^0_{21}} & \underline{\chi^0_{22}} & \cdots & \underline{\chi^0_{2N}} \\ 
\vdots & \vdots & \ddots & \vdots \\
\underline{\chi^0_{N1}} & \underline{\chi^0_{N2}} & \cdots & \underline{\chi^0_{NN}} 
\end{array}\right)
\end{equation} 
and 
\begin{equation}
\underline{\underline{V_{ij}}} = \left(
\begin{array}{cccc}
\underline{V} & \underline{0} & \cdots & \underline{0} \\ 
\underline{0} & \underline{V} & \cdots & \underline{0} \\ 
\vdots & \vdots & \ddots & \vdots \\
\underline{0} & \underline{0} & \cdots & \underline{V} 
\end{array}\right) .
\end{equation} 

The RPA resummation can then be written as
\begin{displaymath}
\underline{\underline{\chi}}=\underline{\underline{\chi^0}}+\underline{\underline{\chi^0}}\; 
\underline{\underline{V}}
\;
\underline{\underline{\chi}}\; 
\end{displaymath}
which is solved by
\begin{equation}
\underline{\underline{\chi}}=\left\lbrack \underline{\underline{1}} -
\underline{\underline{\chi^0}}\; \underline{\underline{V}} \right\rbrack^{-1}
\underline{\underline{\chi^0}} .
\end{equation}

Note that  in this paper we will focus on static correlations.
Since at Gaussian level the coupling between the phase fluctuations and 
the density/amplitude ones is proportional to the frequency\cite{benf04}, 
they are decoupled in the static limit. 
On the other hand, the phase fluctuations enter in a crucial way in the 
calculation of the current fluctuations, as will be outlined in the next subsection.

\subsection{Current Correlations}
The current response $J^{\alpha}_n(\omega)$ to a vector potential $A_x(n,\omega)$ (which we fix along the
 $x$-direction of our square lattice) is 
the sum of the diamagnetic and paramagnetic contribution\cite{scala}
\begin{equation} \label{eq:ja}
J^{\alpha}_n = \sum_m \left\lbrack \delta_{\alpha,x}\delta_{n,m} t_x(n) 
+\chi(j^\alpha_n,j^x_m)\right\rbrack A_x(m)
\end{equation}
where $t_x(n)=-t\sum_\sigma\langle c_{n,\sigma}^\dagger c_{n+x\sigma}
+ c_{n+x,\sigma}^\dagger c_{n\sigma}\rangle < 0$ denotes the kinetic energy 
on the bond
between sites $R_n$ and $R_n+a_x$ and  $j_n^\alpha=-it\sum_\sigma (c^\dagger_{n+\alpha,\sigma}c_{n\sigma}-h.c.)$ is the operator of the paramagnetic current flowing from site $R_n$ to $R_{n+\alpha}$. 
Note that the notation for the current correlation function 
$\chi(j^\alpha_n,j^\beta_m)$
is slightly different from the correlations defined in the previous subsection.
At frequency $\omega=0$ the current only couples to phase
fluctuations $\delta \Phi_i\equiv i(\delta\eta_i-\delta\eta^\dagger_i)/\sqrt{2}$ via the vertices
$\Lambda_{nm}^\alpha=\chi^0(j^{\alpha}_n,\delta\Phi_m)$ 
and $\overline\Lambda_{nm}^\alpha=\chi^0(\delta\Phi_n,j^{\alpha}_m)$. Thus, the full (gauge invariant) current correlation function is then obtained from
\begin{eqnarray}
\chi(j^\alpha_n,j^\beta_m)
&=& \chi^0(j^\alpha_n,j^\beta_m)\nonumber \\
\label{full}
 &+& \Lambda_{nm}^\alpha V_{mk}
\left\lbrack \underline{\underline{1}} - \underline{\underline{\chi^0}}\underline{\underline{V}}\right\rbrack^{-1}_{kl}\overline{\Lambda}_{lm}^\beta ,
\end{eqnarray}
with $\chi^0$ in the second term denoting the bare phase-phase correlation function and
$V_{mk}=-U\delta_{mk}$.

For the Fourier transform of the configurational
average one finally obtains
\begin{equation}\label{eq:jd}
J^{\alpha}_{\bf q} = -D^{\alpha,x}_{\bf q} A_x({\bf q})
\end{equation}
where $D^{\alpha,x}_{\bf q} = -\langle T_x\rangle\delta_{\alpha,x} - 
\langle \chi_{\bf  q}(j^\alpha,j^x)\rangle$. 
For $J^{\alpha}_{\bf q}\equiv J^{x}_{\bf q}$ and taking ${\bf q}$ along the $y$ direction 
the limit $\lim_{q_y\rightarrow 0 } D^{xx}_{q_y}\equiv D_s $ corresponds to the 
superfluid stiffness and 
coincides with the quantity evaluated in Ref.~\onlinecite{sei12}
from an expansion of the mean-field free energy up to quadratic order in
the vector potential. 

\begin{figure}[tbh]
\includegraphics[width=8cm,clip=true]{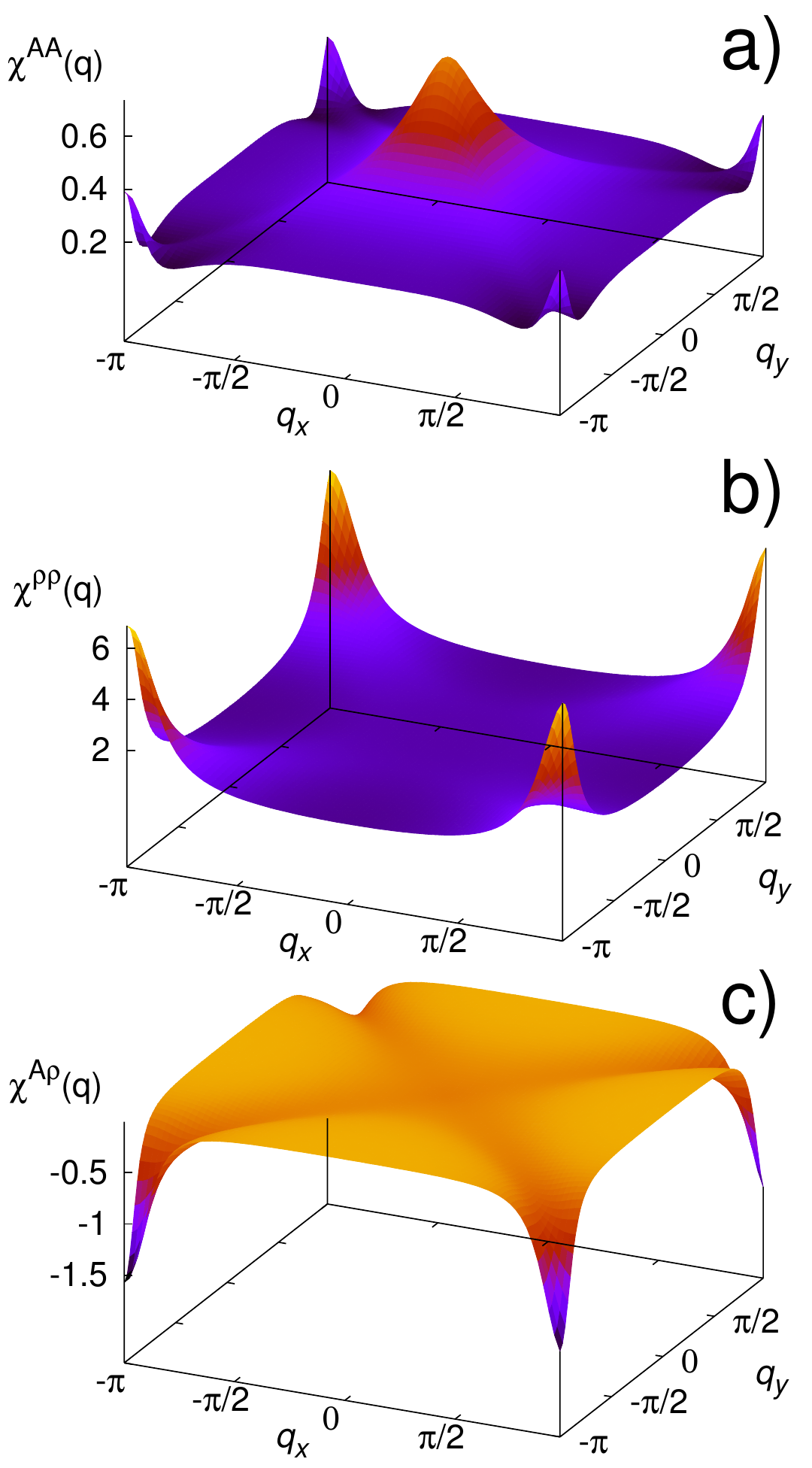}
\caption{(Color online) Top to bottom: amplitude ($\chi^{AA}({\bf q})$), density ($\chi^{\rho\rho}({\bf q})$), and off-diagonal ($\chi^{A \rho}({\bf q})$) 
correlation functions in the superconducting state for parameter $|U|/t=2$ and in the clean limit ($V_0/t=0$).}
\label{fig1}
\end{figure}

\section{Results}\label{sec:res}
\subsection{Correlations in the homogeneous system}

We start our considerations by a brief resume of the homogeneous
case for which amplitude and density correlations have 
been analyzed in Ref.~\onlinecite{lara}
and which are in agreement with our following 
finite cluster analysis.
Fig.~\ref{fig1} shows the amplitude $\chi^{AA}({\bf q})$, density 
$\chi^{\rho\rho}({\bf q})$ and mixed $\chi^{A,\rho}({\bf q})$ 
correlation function for filling
$n=0.875$ and $|U|/t=2$ without disorder.

For these parameters the maximum of the amplitude correlations 
is at ${\bf q}=0$ where it can be approximated as
\begin{equation}\label{eq:fitsmall}
\chi^{AA}({\bf q})\approx \frac{1}{m^2+c q^2}
\end{equation}
with the mass $m$ and a parameter $c$ characterizing
the dispersion of excitations. The quantity $\xi_0=\sqrt{c/m^2}$
can then be interpreted as a length scale for the decay of the
amplitude correlations.
On the other hand the density response is dominated by the contribution
at ${\bf q}={\bf Q}\equiv (\pi,\pi)$ and around this wave-vector can
be described by
\begin{equation}\label{eq:fitq}
\chi^{\rho\rho}({\bf q}\approx {\bf Q})\approx 
\frac{1}{m_Q^2+c_Q ({\bf q}-{\bf Q})^2}\,.
\end{equation}
In real space this corresponds to a staggered decay of the density 
correlations with length scale $\xi_Q =\sqrt{c_Q/m^2_Q}$.

The mixed susceptibility $\chi^{A,\rho}({\bf q})$ is negative (positive)
for densities $n<1$ ($n>1$) since
the anomalous correlations $\langle c_{i\downarrow} c_{i\uparrow}\rangle$
are negative with a maximum of their absolute value at half-filling.
Therefore a positive fluctuation in density $\delta\rho$ for $n<1$ will
lower (i.e. enhance the magnitude) the anomalous correlations.

\begin{figure}[tb]
\includegraphics[width=8cm,clip=true]{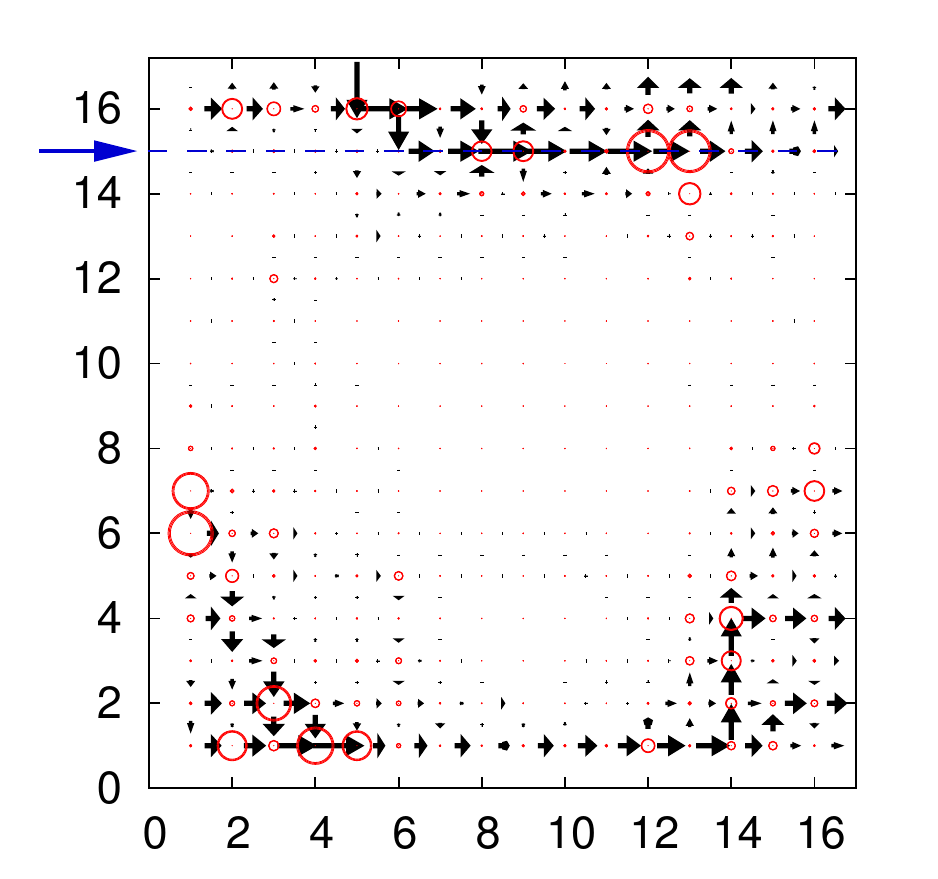}
\caption{(Color online) Distribution of the superconducting gap parameter
$\Delta_i$ (displayed on a linear scale by circles) 
and superconducting currents (arrows) computed
from Eq. (\ref{eq:ja}) for constant vector potential $A_x$ and
a specific disorder configuration. Parameters $|U|/t=5$, $V_0/t=2$.
The dashed line (blue arrow) indicates the cut which is analyzed in
Fig. \ref{fig:cutchi}. }
\label{fig2}
\end{figure}

For the present model, in the absence of disorder and at half-filling,
there is an ``accidental'' symmetry\cite{yan90} which allows the 
superconducting order to be continuously rotated into the 
charge density wave (CDW) order at  ${\bf
    q}={\bf Q}$ without energy change, promoting the charge density mode 
to a Goldstone mode. 
The enhancement of $\chi^{\rho\rho}({\bf Q})$ at $n=0.875$ is a remainder of
this CDW instability at half-filling which is transfered to the amplitude 
correlations via the mixed susceptibility $\chi^{A,\rho}({\bf q})$ 
shown in the bottom panel of Fig.~\ref{fig1}.
Increasing $|U|/t$ enhances the CDW correlations so that at some
point the ${\bf q}={\bf Q}$ amplitude correlations also dominate  with
respect to the ${\bf q}=0$ response. 
On the other hand the CDW correlations are suppressed in the dilute
limit (not shown) so that upon reducing filling the maximum density
response is first shifted away from ${\bf Q}$  along the Brillouin
zone boundary and finally, below some concentration and 
depending on the value of $|U|/t$, the  ${\bf q}=0$ response starts
to dominate . A more detailed
discussion on the filling dependence of the amplitude and
density response in the clean case can be found in Ref.~\onlinecite{lara}.

\subsection{Disordered system: Real space analysis}\label{sec:decoup}
\subsubsection{Mean-field solution}
\label{sec:mean-field-solution}

For sizeable disorder the density varies
on the scale of the lattice constant and correlates with the strongly spatially
fluctuating disorder potential.
Further on, it has been shown in Refs.~[\onlinecite{ghosal98,ghosal01,dubi2007,sei12,ghosh13}] 
that for strong disorder the system disaggregates into SC islands 
with sizeable SC gap $\Delta_i$ which are embedded in regions 
with $\Delta_i\approx 0$. Fig.~\ref{fig2} shows a map of the order
parameter encoded on the size of the red circles showing the formation
of the superconducting islands.  
This island structure leads to a very weak superfluid stiffness. Indeed,
upon applying a transverse vector potential, 
as done in Ref.~\onlinecite{sei12}, the
  current flows through an optimum percolative path or
  ``superconducting backbone'' which determines the global
  stiffness. The latter not only depends on the volume fraction 
  of the superconducting island, but also on the connectivity of these
  islands to  the superconducting backbone. Thus one may
  have a moderate superconducting fraction and a very small global
  stiffness if the connectivity is poor. Fig.~\ref{fig2}
  shows an example of the superconducting backbone for current
  circulation. Notice that it does not  necessarily involve all
  significantly superconducting sites.  For example,  
  sites $(1,7)$ and
  $(3,12)$ in Fig. \ref{fig2}, where $\Delta_i$ is large,  are left out
  which therefore are examples for poorly connected 
  islands. Whereas connected 
  islands determine the superfluid stiffness the disconnected islands dominantly   contribute to the subgap absortpion in the optical conductivity.~\cite{cea14}

Analyzing the
  mean-field solutions for several  configurations of disorder we find that 
there is a strong tendency  to form superconducting dimers. 
For example, for $V_0/t= 2\sim 4$ we
   find that the average number of strongly superconducting neighbors
   of a strongly superconducting site is in the range $0.7 \sim
   0.8$.
Here a strongly superconducting site is defined as a site with a local 
parameter $\Delta_i\geq 0.5\Delta_{max}$ where $\Delta_{max}$ is the largest 
value of  $\Delta_i$ in the  system (which is close to the
maximal value $\Delta_{max}=|U|/2$) . Examples of dimers can be seen in 
Fig.~\ref{fig2} at sites $(1,6)-(1,7)$,  $(12,15)-(13,15)$ and $(8,15)-(9,15)$.
One also observes that dimers can act as seeds of more extended islands
as in sites $(14,3)-(14,4)$.

\begin{figure}[!htp]
\includegraphics[width=7.5cm,clip=true]{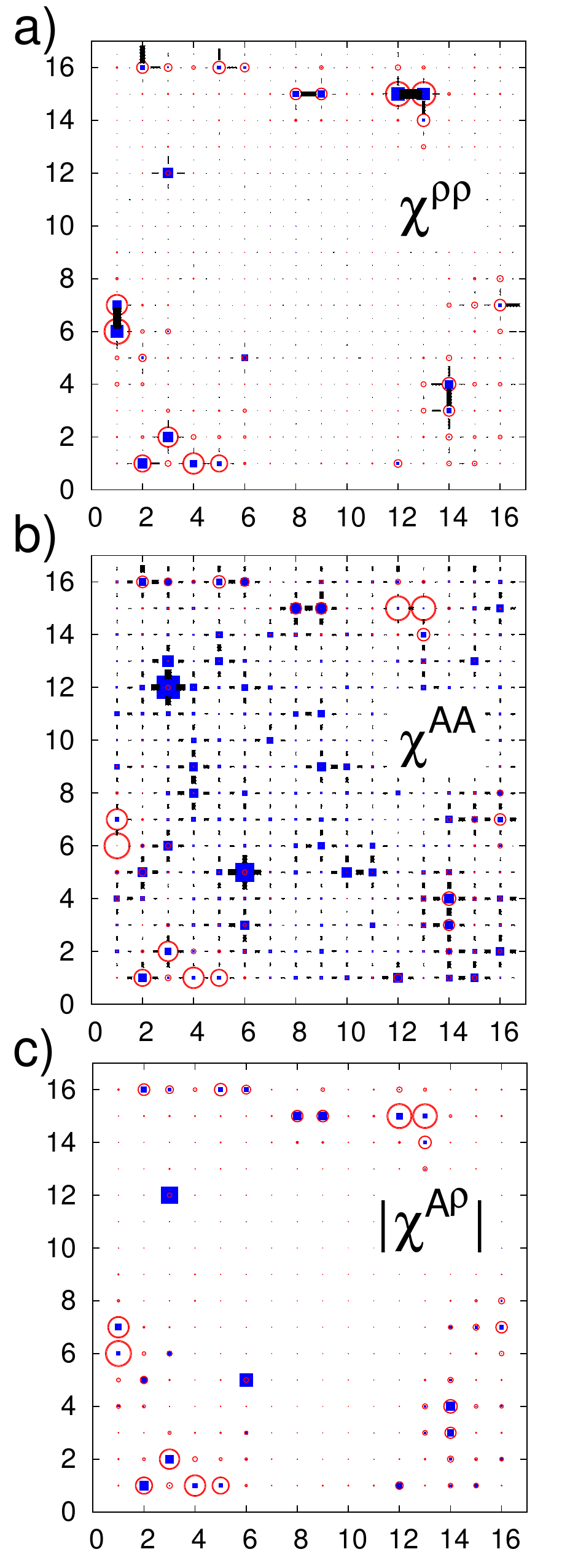}
\caption{(Color online) (a): Distribution of the superconducting gap parameter
$\Delta_i$ (circles), the local density correlation function 
$\chi^{\rho\rho}_{ii}$ (squares), and nearest-neighbor density correlations
$\chi^{\rho\rho}_{\langle ij\rangle}$ (bars on the bonds). 
(b): The distribution of the local $\chi^{AA}_{ii}$ (squares)
and nearest-neighbor $\chi^{AA}_{\langle ij\rangle}$ 
(bars on the bonds)
amplitude correlations 
 together with the SC gap (circles). 
(c): Magnitude of local off-diagonal amplitude-density correlations
$|\chi^{\rho A}_{ii}|$ (squares) together with the SC gap (circles).
The disorder configuration and parameters are the same as in 
Fig.~\ref{fig2}. The symbol size for the correlations is 
displayed on a logarithmic scale whereas the SC gap is plotted on a 
linear scale.}
\label{fig7}
\end{figure}

In previous work ~\cite{ghosal98,ghosal01} it has been found that
the preferable sites for the SC islands are those with 
the Hartree potential $H_i=-|U|\langle n_i\rangle +V_i$
being close to the chemical potential $\mu$, since this allows for strong
particle-hole mixing. This would imply that the 'good' SC sites 
are already encoded in the normal state since there exists
a strong correlation between the local Hartree 
potentials in the SC and normal state.
On the other hand the correlation between $H_i$ and the
size of $\Delta_i$ weakens with increasing disorder, i.e.
a small $|H_i-\mu|$ not necessarily correlates with a large
$\Delta_i$ whereas a large $\Delta_i$ always implies a
small $|H_i-\mu|$. A similar conclusion has been drawn in 
Ref. \onlinecite{ghosh13} where the relation of order parameter
variations and the shell effect has been investigated.

 \begin{figure}[tb]
\includegraphics[width=8cm,clip=true]{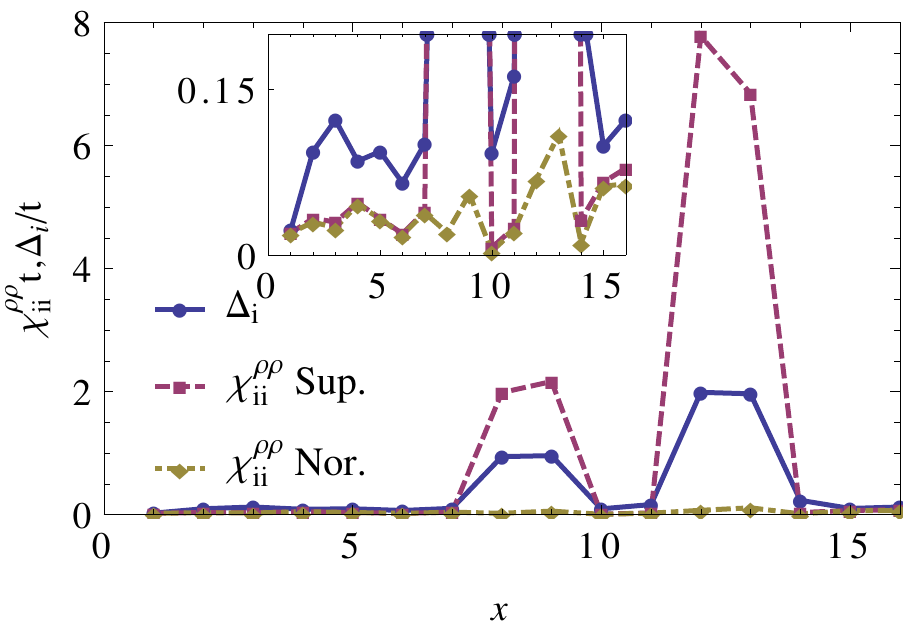}
\caption{(Color online) Cut of the order parameter distribution 
(full line, circles) and local density susceptibility  $\chi_{ii}^{\rho\rho}$ 
in the superconducting state (dashed line,
  squares) and in the normal state (dot dashed line, diamonds). The
  cut is done along the row with $y=15$ of Fig.~\ref{fig7}b and
is indicated by an arrow in Fig.~\ref{fig2}. }
\label{fig:cutchi}
\end{figure}

\subsubsection{Real space structure of responses}\label{secrr}

The largest contribution to the density and amplitude correlations comes 
from the diagonal elements $\chi_{ii}^{AA}$ and $\chi_{ii}^{\rho\rho}$ that 
are shown  
as a  logarithmic map in Fig.~\ref{fig7}a and b, respectively. 
Here the disorder realization is the same as in Fig.~\ref{fig2} and the 
local SC gap is shown with circles, whose size is proportional to the 
gap magnitude. Panel (a) shows also the nearest-neighbor density-density
correlation $\chi^{\rho\rho}_{ij}$ encoded in the size of the bars on
the bonds. 

One finds that the strong superconducting sites coincide with sites
which have a large charge density susceptibility.
Also the dominant nearest neighbor 
density correlations $\chi^{\rho\rho}_{\langle ij\rangle}$ are attached
to the SC islands and become particularly enhanced among the sites
forming a SC dimer.  We find that 
the bare local density correlations $\chi^{0,\rho\rho}_{ii}$ in the SC state
show a similar structure (not shown) but with smaller absolute value 
($\sim 1/20$).

This rises the ``chicken and
egg'' question if sites are favorable for superconductivity 
because they have a large susceptibility already in  the normal state
or if the large susceptibility is due to the local superconducting
correlations. 
To answer this question we have computed the charge density susceptibility in
the absence of superconductivity. Although there is a tendency for
sites with charge density susceptibility larger than the average in the normal
state to become superconducting, there is an enormous enhancement of the charge
density susceptibility on the superconducting sites. This can be seen in the
cut of the local susceptibilities and order parameter shown in
Fig.~\ref{fig:cutchi}. We see that on the superconducting sites
the local susceptibility can be enhanced by two orders of
magnitude. The inset shows a zoom of the intensity scale showing that
the superconducting sites tend to have a charge density susceptibility larger
than the average in the normal state but which does not explain the
enhancement seen in the superconducting state. It also shows that on
the sites with small order parameter the charge susceptibility remains
the same in the superconducting and normal state. Clearly this
behavior is due to the almost incompressible character of the phase
without superconducting correlations 
which becomes instead highly compressible in the superconducting
state. This physics is similar to that in the clean half-filled
Hubbard model where a rotation between the two competing states, 
CDW and SC, essentially induces a transition from zero
to very large compressibility $\kappa$.

\begin{figure}[tbh]
\includegraphics[width=8cm,clip=true]{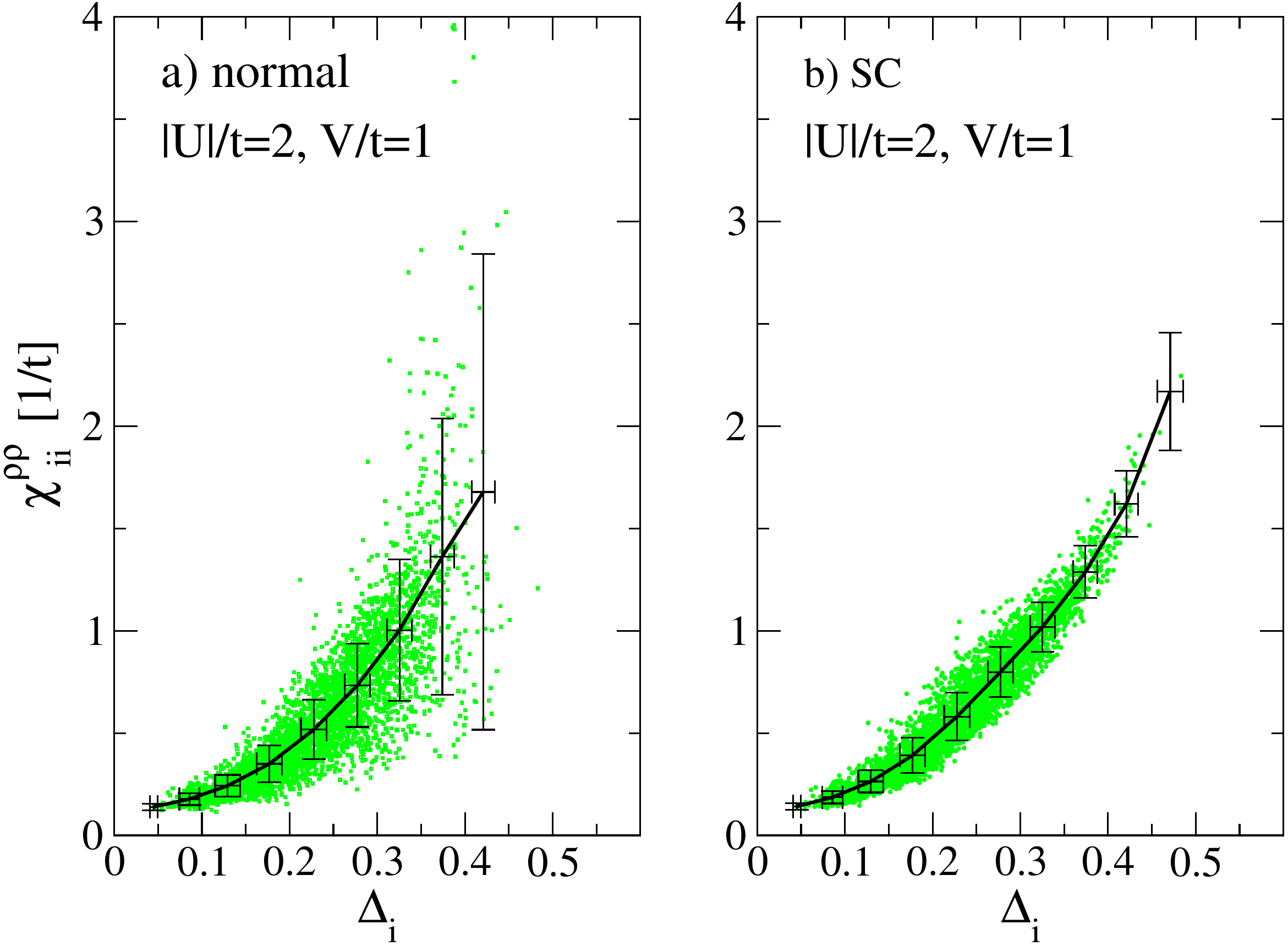}
\includegraphics[width=8cm,clip=true]{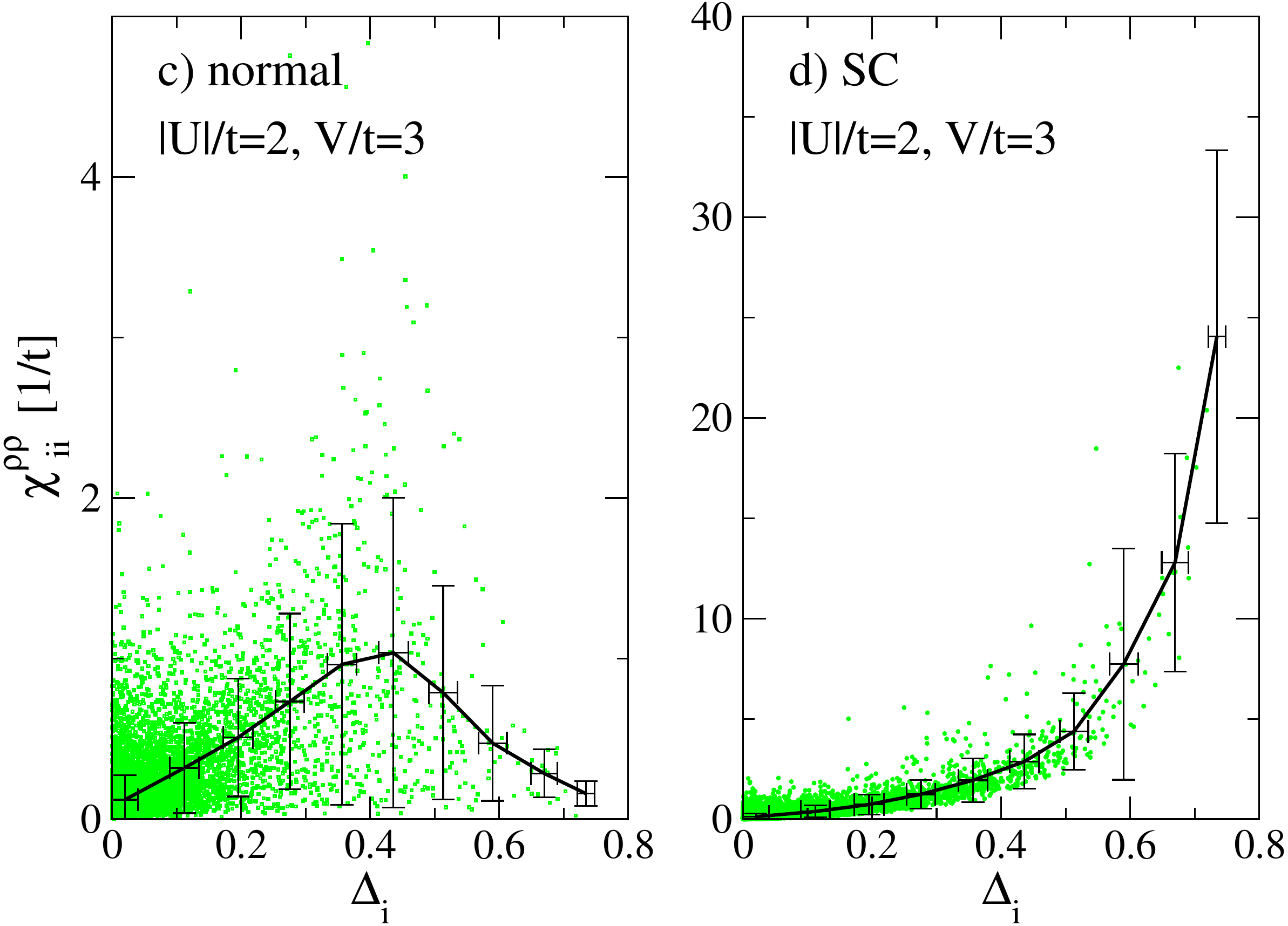}
\caption{(Color online) Plot of the points $(\Delta_i,\chi_{ii}^{\rho\rho})$ (green) for the
normal (left panels a, c) and SC (right panels b, d) system
where $\Delta_i$ refers to the value in the SC state. 
The lines and errorbars have been
obtained by collecting data in  $10$ bins of $\Delta$. 
$|U|/t=2$, $V/t=1$ (upper panels a, b), $V/t=3$ (lower panels c, d).}
\label{corrgap}
\end{figure}

The correlation between SC gap and local charge density susceptibility is summarized
in Fig. \ref{corrgap} which shows the distribution of 
$(\Delta_i, \chi_{ii}^{\rho\rho})$ points from $200$ samples for the normal and SC state and two
values of disorder at $|U|/t=2$. Here $\Delta_i$ always refers to the
value in the SC state whereas $\chi_{ii}^{\rho\rho}$ is evaluated in both
normal and SC state. In the normal state and 
for weak disorder $V/t=1$ one
observes a positive correlation between the local $\chi_{ii}^{\rho\rho}$
and the gap $\Delta_i$ which would develop in the SC state.   
This correlation gets sharper in the SC state (panel b) but extends
over the same range of $\chi_{ii}^{\rho\rho}$ values than in the normal state.
In contrast, for larger disorder $V/t=3$ there is almost no correlation
between local charge density susceptibility and SC gap in the
normal state while this correlation is strongly enhanced in the SC state
and pushed to values of $\chi_{ii}^{\rho\rho}$ which are one order of magnitude
larger than in the normal state.

\begin{figure}[tb]
\includegraphics[width=8cm,clip=true]{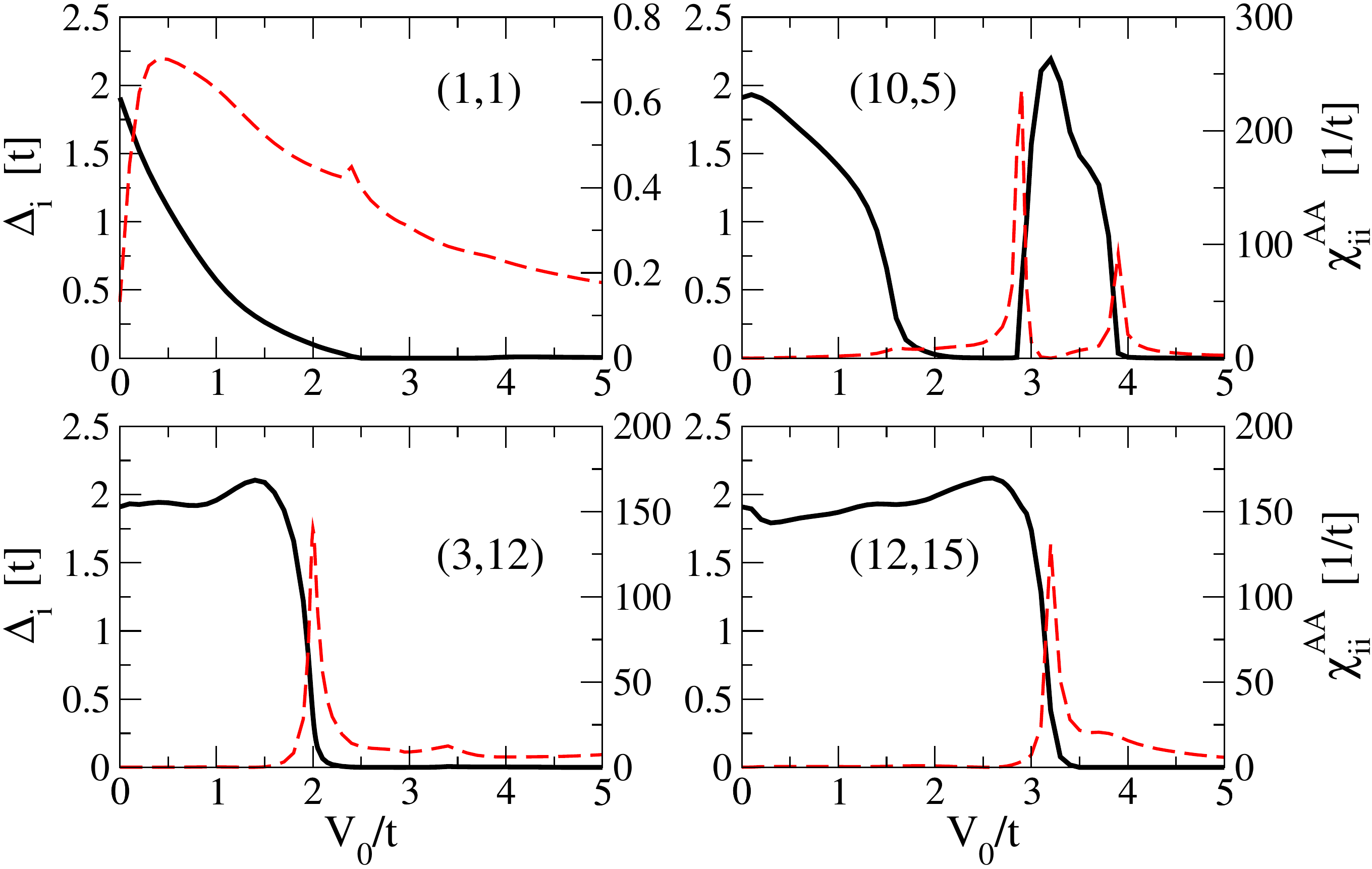}
\caption{(Color online) Disorder dependence of the SC gap (solid, black) and of the local
amplitude correlations (red, dashed) for selected sites of the disorder
configuration used in Figs.~\ref{fig2},\ref{fig7}. $|U|/t=5$.}
\label{fig8}
\end{figure}

The behavior of the amplitude fluctuations is also very interesting. 
We find that local amplitude fluctuations are significantly enhanced when 
the SC gap displays strong variations as a function of disorder strength.
This feature is exemplified in Fig.~\ref{fig8} which, for fixed
disorder realization 
(the same as used in Fig. \ref{fig2} and Fig. \ref{fig7}), shows
the dependence of $\chi_{ii}^{AA}$ on $V_0$ 
for selected sites. One basically observes two kinds of behavior.
First there are 'weak' SC sites, as $(1,1)$ or $(10,5)$, whose order 
parameter immediately decreases with the onset of disorder. Besides there
are 'strong' SC sites, as $(3,12)$ or $(12,15)$ which initially resist disorder
and where $\Delta_i$ can even get enhanced with respect to its $V_0=0$ value. 
The drop of $\Delta_i$ on the strong SC sites at a given $V_0/t$ is then
accompanied by a peak
in $\chi_{ii}^{AA}$ resembling the behavior close to a
second order phase transition. However, the order parameter
does not vanish on the disordered site of the transition but acquires
a small finite value due to the proximity effect of other SC islands.
For a given disorder strength
only few sites are close to this regime and 
their number decreases with increasing $V_0/t$ due to the decrease of
SC islands.
There are also few sites, as $(10,5)$, where the SC order 
parameter reemerges at a large value of the disorder strength
and stays finite over some range of $V_0$. In the appendix
the behavior of $\Delta_i$ and $\chi_{ii}^{AA}$ for all
sites of the sample is analyzed in more detail.

The present real space analysis reveals that in the strongly
disordered regime, density correlations are dominant on the SC islands
whereas the amplitude correlations are large in the other part
of the system, i.e. where the SC gap is almost completely suppressed by disorder.
As shown in Fig. \ref{fig7}c there are only few sites 
with significant off-diagonal correlations $\chi_ {ii}^{\rho,A}$.
Besides on the 'marginal' sites $(3,12)$ and $(6,5)$, which are at the
transition $\Delta\to 0$, the mixing
of amplitude and density correlations is only observed on some
of the SC sites. Clearly this decoupling of amplitude and density
correlations will be even more pronounced in the average momentum dependent 
correlations which will be analyzed in the next subsections.

\subsection{Disordered system: Fourier space analysis}
For a particular disorder configuration the Fourier transform
of the correlation functions is given by
\begin{equation}
\chi({\bf q}, {\bf q}')=\frac{1}{N}\sum_{ij}e^{i\left({\bf q}{\bf R}_i
-{\bf q}'{\bf R}_j\right)}\chi_{ij}
\end{equation}
where $N$ denotes the number of lattice sites.
Clearly, if $\chi_{ij}$ only depends on the distance between lattice
sites ${\bf R}_i-{\bf R}_j$ then $\chi({\bf q}, {\bf q}')$ is
diagonal in momenta. In the following we perform averages
of $\chi_{ij}$ over different disorder realizations up to $n_d=200$
for lattice sizes up to $24\times 24$.
This procedure restores translational invariance in the correlation
functions so that  $\langle \chi({\bf q}, {\bf q}') \rangle_{conf.}
\equiv \delta({\bf q}, {\bf q}') \chi({\bf q})$.
In Figs.~\ref{fig6}, \ref{fig4} the errorbars in the compressibility and
mass reflect the variance of $\chi({\bf q})$ at ${\bf q}=0$ and
${\bf q}={\bf Q}$, respectively. Although it increases with disorder
and $|U|/t$ the mean-values exceed the variances for the 'worst' cases
by a factor $\sim 3$.

We fit the correlation function $\chi({\bf q})$, which is peaked at
${\bf q}={\bf Q}$, to the function 

\begin{equation}\label{eq:fitqc}
\chi({\bf q})=\lambda_0+\frac{\lambda_3}{1
+2 \lambda_1 \gamma_1({\bf q}-{\bf Q})
+2 \lambda_2 \gamma_2({\bf q}-{\bf Q})}
\end{equation}
with
\begin{eqnarray*}
\gamma_1({\bf q}) &=& 2-\cos(q_x)-\cos(q_y) \\
\gamma_2({\bf q}) &=& 1-\cos(q_x)\cos(q_y)\,.
\end{eqnarray*}

Although Eq. (\ref{eq:fitqc}) yields a good account of the correlations
over the whole Brillouine zone (BZ)
the fit is restricted to an area of $\approx 5\%$ of the BZ 
around the peak at ${\bf Q}$ in order to extract the parameters
in Eqs. (\ref{eq:fitsmall},\ref{eq:fitq}). Expanding Eq. (\ref{eq:fitqc})
around ${\bf Q}$ yields
\begin{eqnarray}
m^2&=&\frac{1}{\lambda_0+\lambda_3} \label{eq:fitm}\\
c&=&\frac{\lambda_1+\lambda_2}{(\lambda_0+\lambda_3)} \label{eq:fitc}\\
\xi^2&=& c/m^2 = \lambda_1+\lambda_2 \,.\label{eq:fitxi}
\end{eqnarray}

In the following we analyze the momentum structure of the averaged
density-, off-diagonal and amplitude correlations.
The various
  fitting parameters will be distinguished by (a) the reference momentum 
in the expansion,
i.e. ${\bf q}=0$ or ${\bf Q}\equiv (\pi,\pi)$, and (b) a superscript
which indicates the correlation function.
For example, $\xi_0^{A}$  will denote the 
correlation length for amplitude fluctuations derived from an expansion
of $\chi^{AA}({\bf q})$ around ${\bf q}=0$.

\begin{figure}[tbh]
\includegraphics[width=8cm,clip=true]{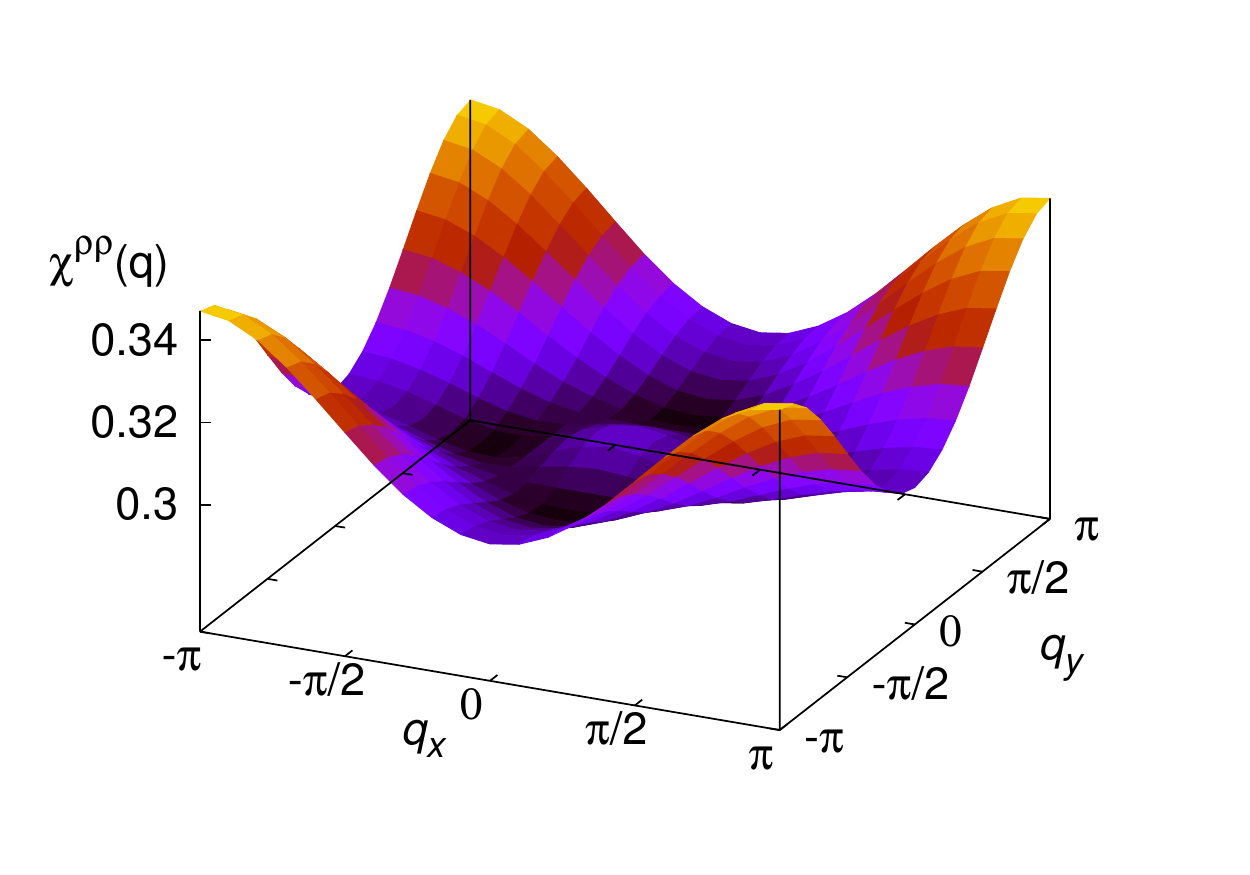}
\caption{(Color online) Average (number of samples $=200$) of Fourier transformed 
density correlations for parameter $|U|/t=2$, $V_0/t=3$.}
\label{fig5}
\end{figure}
 
\subsubsection{Momentum structure of $\chi^{\rho\rho}({\bf q})$}\label{secrr}
We start with the analysis of the momentum dependence of 
the averaged density correlation function which is 
shown in Fig.~\ref{fig5} for parameters $|U|/t=2$ and $V_0/t=3$.
Disorder induces an overall suppression of the response as
compared to the clean case in Fig.~\ref{fig1}b.
This is most pronounced
for the CDW correlations at ${\bf q}={\bf Q}$ which for $V_0/t=3$ are
reduced by a factor $1/20$ with respect to the clean case correlations.
At ${\bf q}=0$ this reduction is only $1/2$ so that in Fig.~\ref{fig5}
one observes a relative enhancement of the zone center correlations.
For $|U|/t=2$ the crossover from dominant CDW to  ${\bf q}=0$ correlations
occurs at $V/t \approx 4$ whereas for larger values ($|U|/t=5$) 
$\chi^{\rho\rho}({\bf q})$ has a minimum at ${\bf q}=0$ up to
the largest disorder investigated. 
Note that also for smaller filling disorder shifts the
dominant correlations from incommensurate momenta in the clean
case to  ${\bf Q}=(\pi,\pi)$ so that the following analysis is representative
for a wide doping range and disorder values.

\begin{figure}[tbh]
\includegraphics[width=8cm,clip=true]{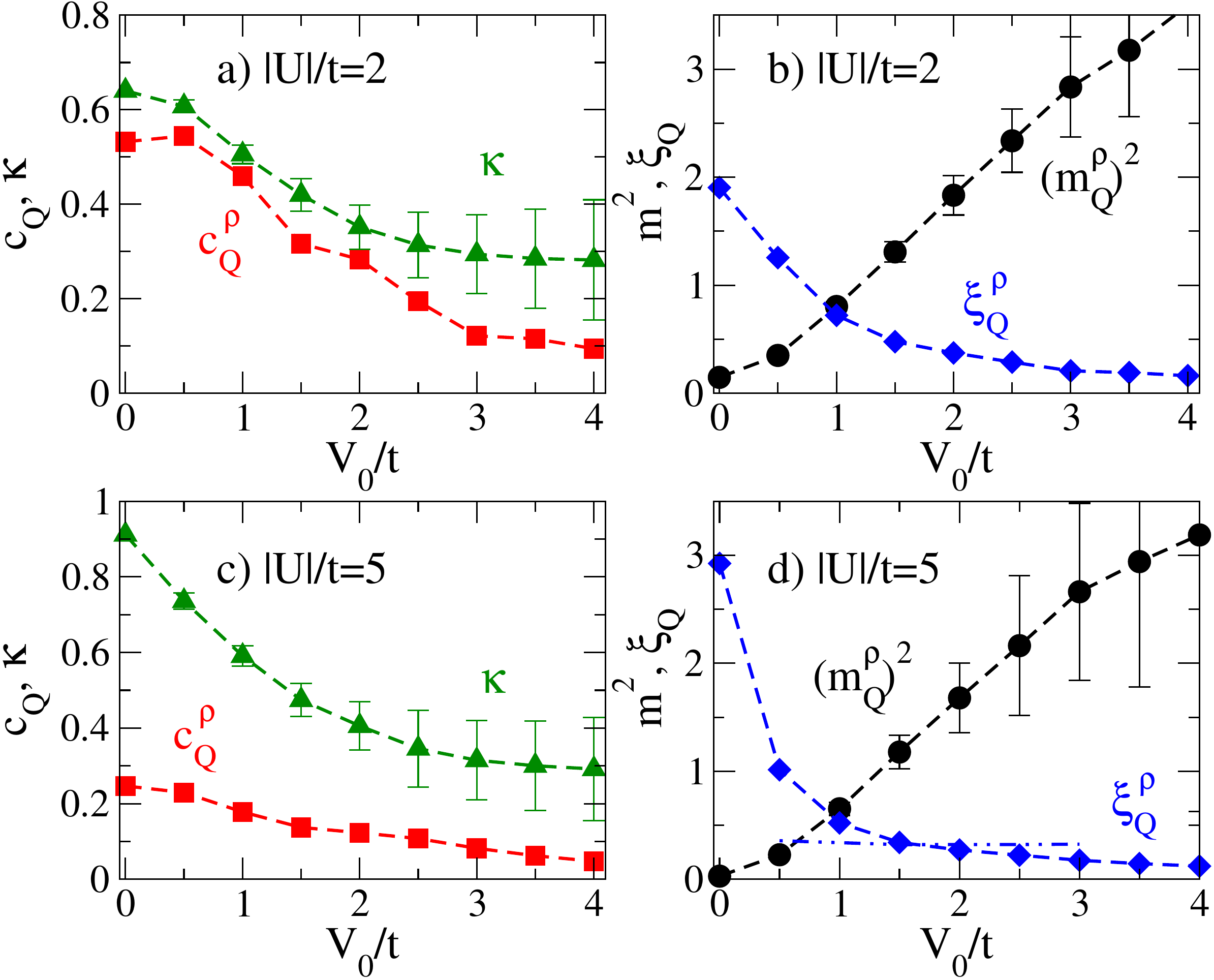}
\caption{(Color online) Disorder dependence of the fit parameters $(m_Q^\rho)^2$ (circles),
$c_Q^\rho$ (squares), and $\xi_Q^\rho$ (diamonds) for the
staggered density correlations
extracted from Eqs. (\ref{eq:fitqc} - \ref{eq:fitxi}).
The disorder dependence of the compressibility is shown by the triangles.
The dashed-dotted line in panel d) indicates the correlation length in the
normal state. In panel b) the normal state $\xi_Q^\rho$ is numerically 
identical to the result in the SC state.}
\label{fig6}
\end{figure}

Fig.~\ref{fig6} shows the parameters $(m_Q^\rho)^2$ and $c_Q^\rho$
obtained from the fit to Eq.~\ref{eq:fitqc} with ${\bf Q}=(\pi,\pi)$ 
as a function of disorder together with the 
compressibility $\kappa=\chi^{\rho\rho}_{{\bf q}=0}$.

In the strong coupling limit (small $2t/|U|$) the clean case compressibility
scales as  $\kappa\approx |U|/8t^2$.~\cite{lara} The enhancement of
$\kappa$ with $|U|/t$ can also be observed in Fig.~\ref{fig6} for
$V_0/t=0$ although the parameters $|U|/t=2,5$ are rather in the
intermediate coupling regime so that the agreement with the
above estimate is only qualitative.
Upon increasing $V_0/t$  there is first a decrease of $\kappa$, in agreement 
with the results of Refs. \onlinecite{ghosal98,ghosal01}. 
At large disorder one observes a tendency 
of the average compressibility $\kappa$ to saturate to a value that is weakly 
dependent on $U$. 
Since in this regime
the dominant contribution to $\kappa$ comes from the (real space) diagonal
elements $\chi_{ii}^{\rho\rho}$ on the SC islands there exists an
apparent inverse correlation between the number of SC islands (which
decreases with $V_0/t$) and the local compressibility $\chi_{ii}^{\rho\rho}$
(which gets enhanced with increasing $V_0/t$).

We now turn to the analysis of the CDW correlation length in the disordered
SC system. For weak disorder $V_0/t=0.5$ there is a strong difference
in the density distribution obtained for the 
two values of $|U|/t=2, 5$ which we have investigated. In fact,
for $|U|/t=2$ we find that  
the difference in the density distribution between normal
and SC state is small for each value of the disorder potential $V_0/t$.
As a consequence the decrease of the CDW correlation length
with $V_0/t$ (Fig. \ref{fig6}b) is the same in the normal
and SC state within the numerical accuracy.
On the other hand, for $|U|/t=5$ we find that already for 
$V_0/t=0.5$ sites in the normal state system are either
almost empty or doubly occupied. As already discussed above, the SC state 
induces a redistribution of charge density which in this case leads to a 
significant rearrangement 
with a more homogeneous distribution between $n \approx 0.2$ and $n\approx 1.7$.
As a consequence of this effectively less disordered SC state one 
observes in panel (d) of Fig. \ref{fig6}
an enhancement of the correlation length at $V_0/t=0.5$ 
from $\xi_Q^\rho\approx 0.3$ in the normal state to $\xi_Q^\rho\approx 1$ 
in the SC system.
 
The behavior of fit parameters in the SC system, 
as shown in Fig. \ref{fig6}, can then be qualitatively 
understood from the evolution toward the bimodal charge density distribution, 
where the low (high) density peak
approaches $n_L=0$ ($n_H=2$) 
with increasing disorder.
We also adopt the result from 
a strong coupling expansion of $\chi^{\rho\rho}({\bf q})$
for the homogeneous system \cite{lara,note1} 
which for the  mass parameter yields
\begin{equation}\label{m2sc}
m_Q^2=\frac{8t^2}{|U|}\frac{\delta^2}{1-\delta^2}
\end{equation}
and $\delta=1-n$ denotes the doping measured from half-filling.
Averaging Eq.~\ref{m2sc} over the bimodal distribution. 
yields $\langle m_Q^2\rangle
\sim (\delta n)^2/(1-(\delta n)^2)$ with $\delta n =n_H - n_L$.
The grow of $\delta n$ with $V_0/t$ then 
accounts for the increase of $m_Q^2$ with disorder as shown in
Fig.~\ref{fig6}.

In the strong-coupling clean case the parameter $c_Q$ is given by \cite{lara}
\begin{equation}
c_Q=\frac{t^2}{|U|}\frac{1-2\delta^2}{1-\delta^2}=\frac{t^2}{|U|}-\frac{m_Q^2}{8}
\end{equation}
and is thus expected to decrease with disorder proportional to
the increase of $m_Q^2$. Within the numerical error this is in fact the
behavior observed in Fig.~\ref{fig6} and also accounts for the
decrease of the correlation length $\xi_Q$ with disorder.

\subsubsection{Momentum structure of amplitude correlations}\label{seccaa}
\begin{figure}[tbh]
\includegraphics[width=8cm,clip=true]{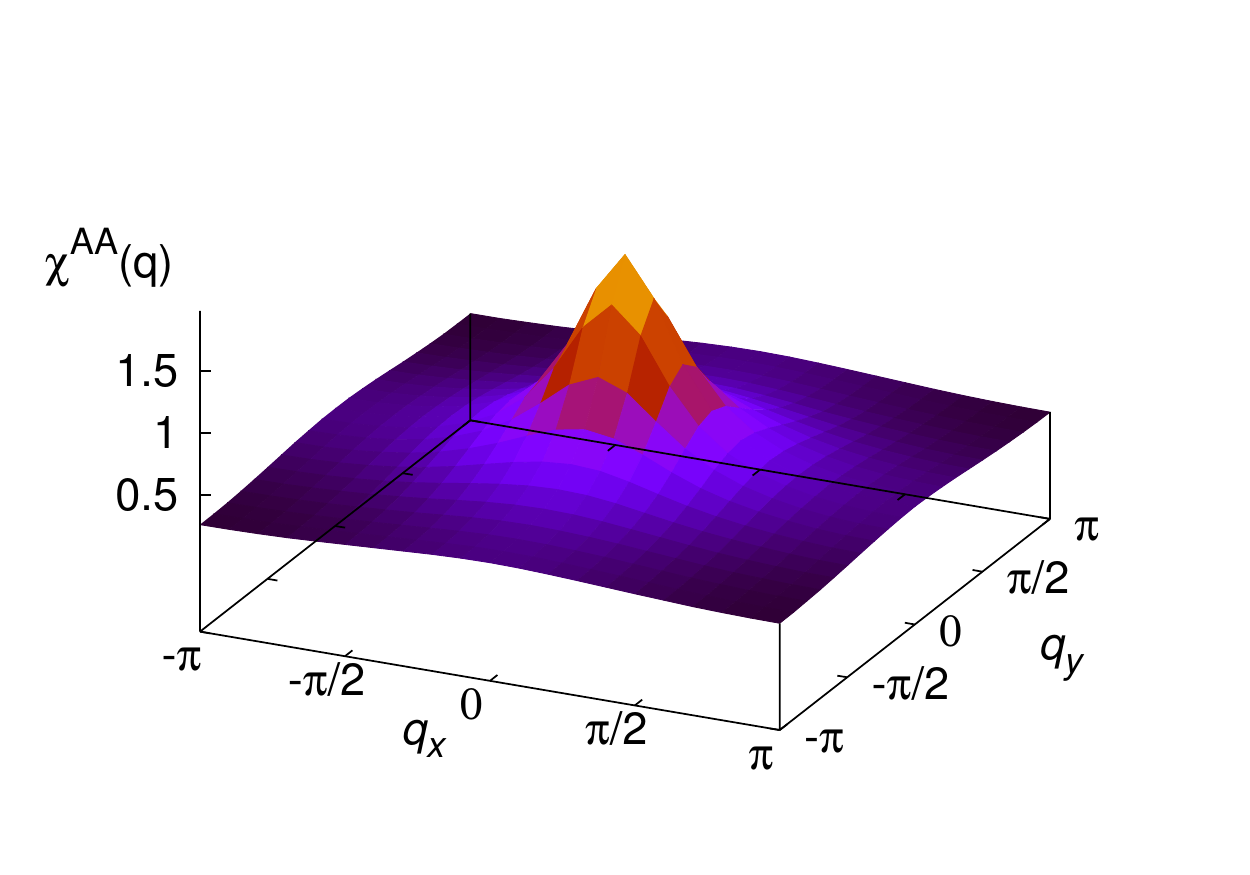}
\caption{(Color online) Average (number of samples $=200$) of Fourier transformed 
amplitude correlations for parameter $|U|/t=2$, $V_0/t=3$.}
\label{fig3}
\end{figure}

We proceed by analyzing the amplitude correlations 
$\chi^{AA}({\bf q})$ on top of
the BdG solution whose momentum dependence is reported in
Fig.~\ref{fig3} for $|U|/t=2$, $V_0/t=3$.

It turns out that disorder removes the enhancement
of amplitude correlations at  
${\bf Q}=(\pi,\pi)$, which were dominating in the clean case
for this value of $|U|/t$. An interesting result is the
concomitant enhancement of the ${\bf q}=0$ response by a
factor of $\sim 5/2$ which therefore dominates the 
amplitude correlations for large disorder.
As we have seen in the previous section, the density correlations are still
peaked at ${\bf Q}=(\pi,\pi)$ for these parameters 
 which indicates the decoupling of
density and amplitude fluctuations with increasing disorder.
Note that in contrast to the density correlations,
the amplitude fluctuations in the normal state will always
be unstable.

\begin{figure}[tbh]
\includegraphics[width=8cm,clip=true]{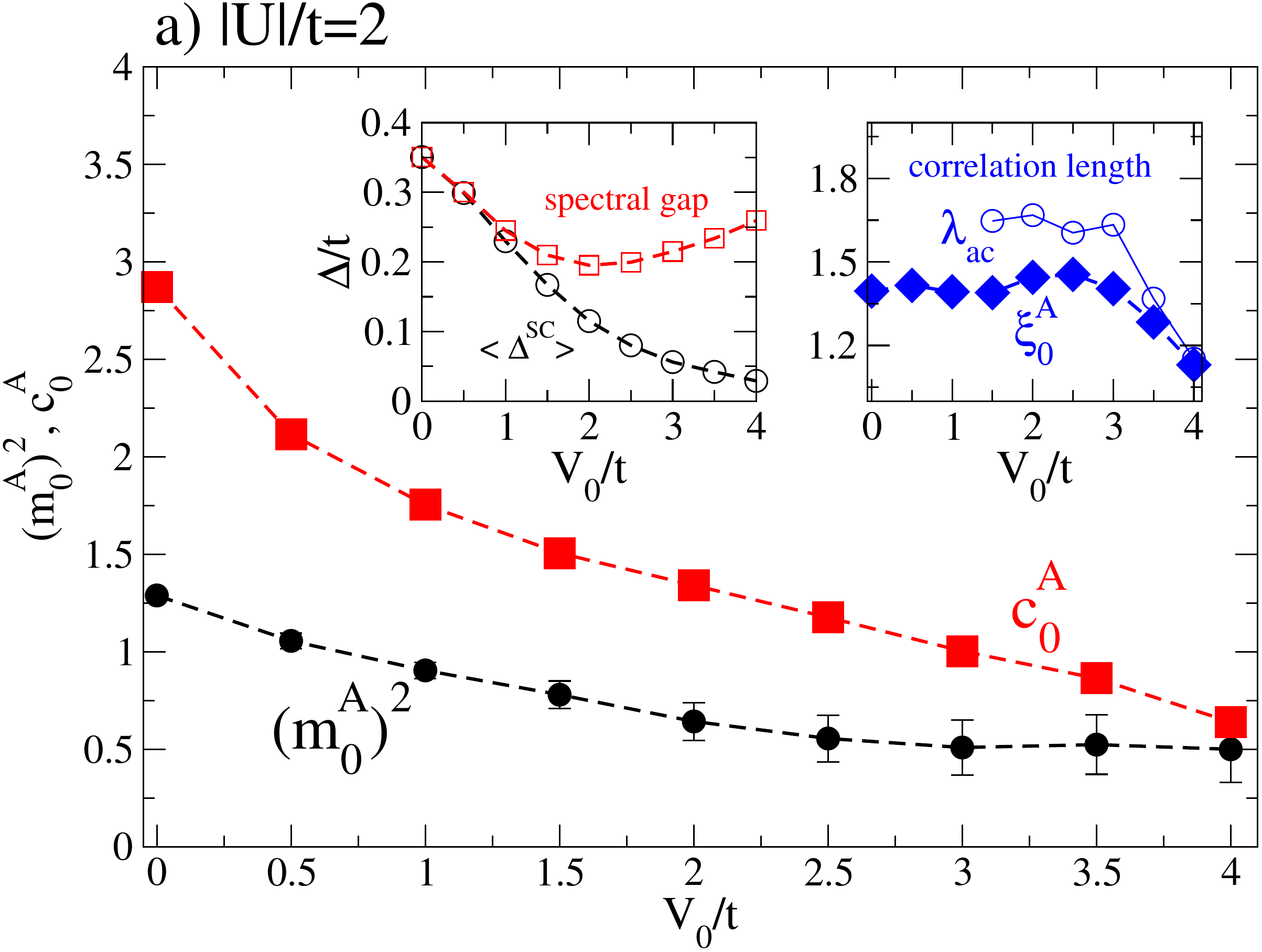}
\includegraphics[width=8cm,clip=true]{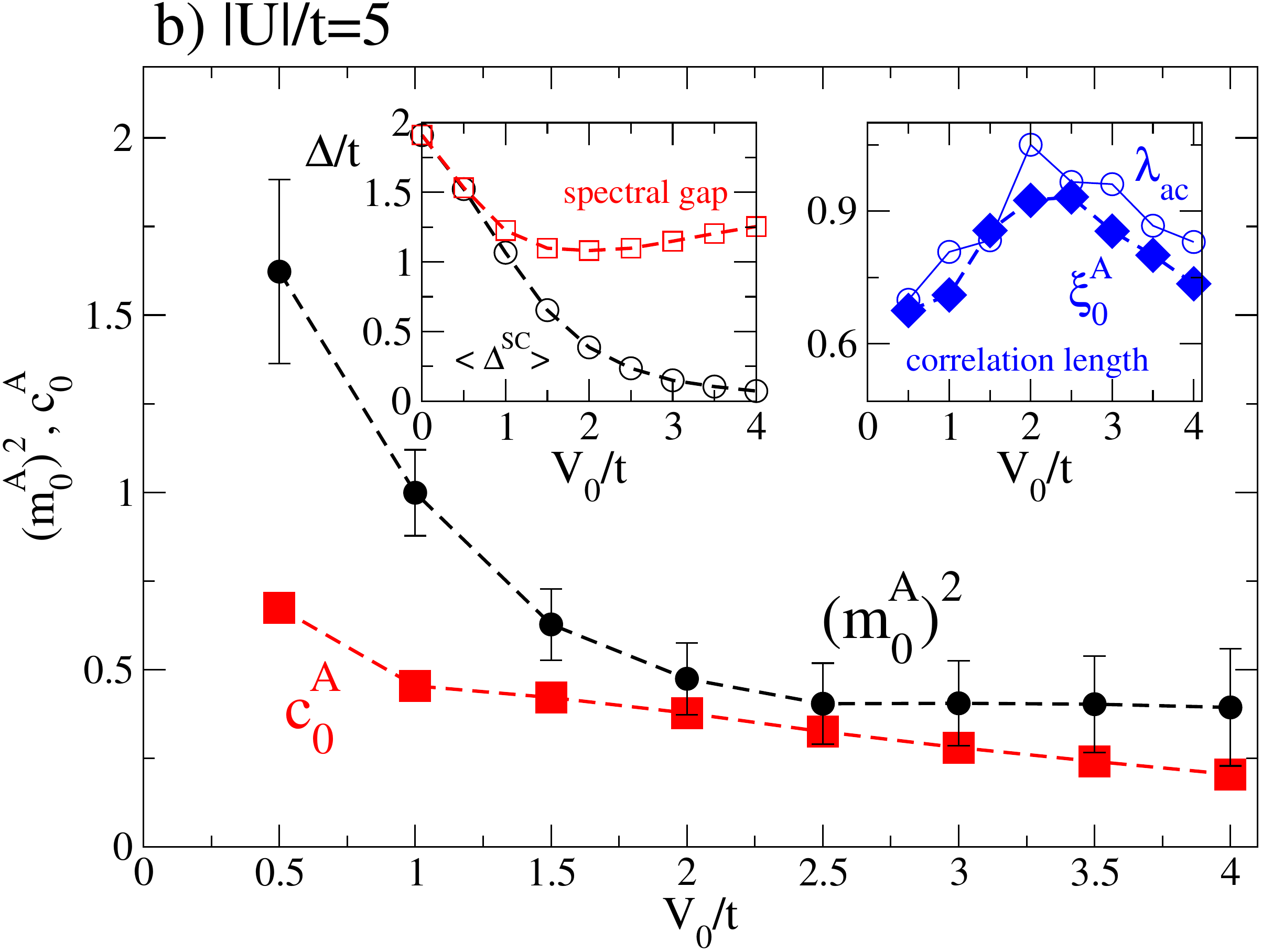}
\caption{(Color online) Disorder dependence of the fit parameters $(m0^A)^2$ (circles),
$c_0^A$ (squares) and $\xi_0^A=\sqrt{c_0^A/(m_0^A)^2}$ (diamonds, right inset) as
extracted from Eqs. (\ref{eq:fitqc} - \ref{eq:fitxi}) for
$|U|/t=2$ (a) and $|U|/t=5$ (b). 
The left inset reports the average superconducing gap (circles)
and average spectral gap (squares). The right insets
also show the gap autocorrelation length $\lambda_{ac}$ (circles) 
computed from Eqs. 
(\ref{eq:autocorr},\ref{fit2d}).}
\label{fig4}
\end{figure}

The latter are again characterized by the
mass $(m_0^A)²$ and $c_0^A$  parameter
obtained from the fit of $\chi^{AA}({\bf q})$
to Eq.~\ref{eq:fitqc} around ${\bf q}=(0,0)$.
Fig.~\ref{fig4} reports the fit parameters as a function of disorder,
again for values of the onsite attraction $|U|/t=2$ and $|U|/t=5$.
Note that for the larger interaction $|U|/t=5$ and small disorder the
correlations show the dominant peak at ${\bf Q}=(\pi,\pi)$ 
for which reason the fit parameters
are only reported for $V_0/t \ge 0.5$.

The aforementioned enhancement of the ${\bf q}=(0,0)$ amplitude 
correlations with $V_0/t$ now results in the decrease of
the mass $m_0^A$  with disorder with tendency to saturate at large 
$V_0/t \gtrsim 2$. Also the parameter $c$ decreases with 
the disorder strength so that the resulting correlation length 
$\xi_0^A=c_0^A/(m_0^A)^2$ (right insets to Fig.~\ref{fig4})
crucially depends on the relative change of $c_0^A$ and $(m_0^A)^2$
with $V_0/t$.

For $|U|/t=2$ the correlation length is almost constant up to
$V_0/t=2.5$ and then starts to decrease with disorder.
For larger $|U|/t$ one even observes an enhancement for
small $V_0/t$ so that $\xi_0^A$ acquires a maximum around
$V_0/t=2.5$. We note that this is not an effect of
competing CDW order since the same result is observed
in the low-density regime where such correlations are absent.

In the limit of small $V_0/t$ one can adopt the usual expression
for the correlation length in dirty superconductors given by
$\xi_0=\sqrt{\xi_{BCS} l}$ with the mean free path $l$ and
the correlation length of the clean system $\xi_{BCS} \sim v_F/\Delta^{SC}$.
The behavior of $\xi_0(V_0)$ therefore crucially depends on the
depletion of the density of states, which lowers the superconducting
$\Delta^{SC}$ gap, and the reduction of the mean free path $l$ with disorder.
As noted in Ref.~\onlinecite{ghosal98,ghosal01} the situation in the
strongly disordered system is more interesting since one has
to distinguish between the average superconducting order parameter
$\langle\Delta^{SC}\rangle$ and the spectral gap. As shown in the
left insets to Fig.~\ref{fig4} $\langle\Delta^{SC}\rangle$ continuously
decreases with disorder due to the increase of the 'non-SC' area.
On the other hand the spectral gap first shrinks with
disorder due to the depletion of the density of states but grows
again for strong disorder, signaling the formation of local boson pairs 
that get progressively localised as the SIT is approached. 
One can then argue that at strong
disorder the BCS correlation length tends to scale as the inverse of the 
spectral gap, that acts as a cut-off to the increase of $\xi_0$ associated 
to the suppression of the SC order parameter. 
Alternatively one can relate the disorder dependence of the
correlation length to the behavior of the nearest-neighbor amplitude
correlations as shown in the appendix.

\begin{figure}[thb]
\includegraphics[width=8cm,clip=true]{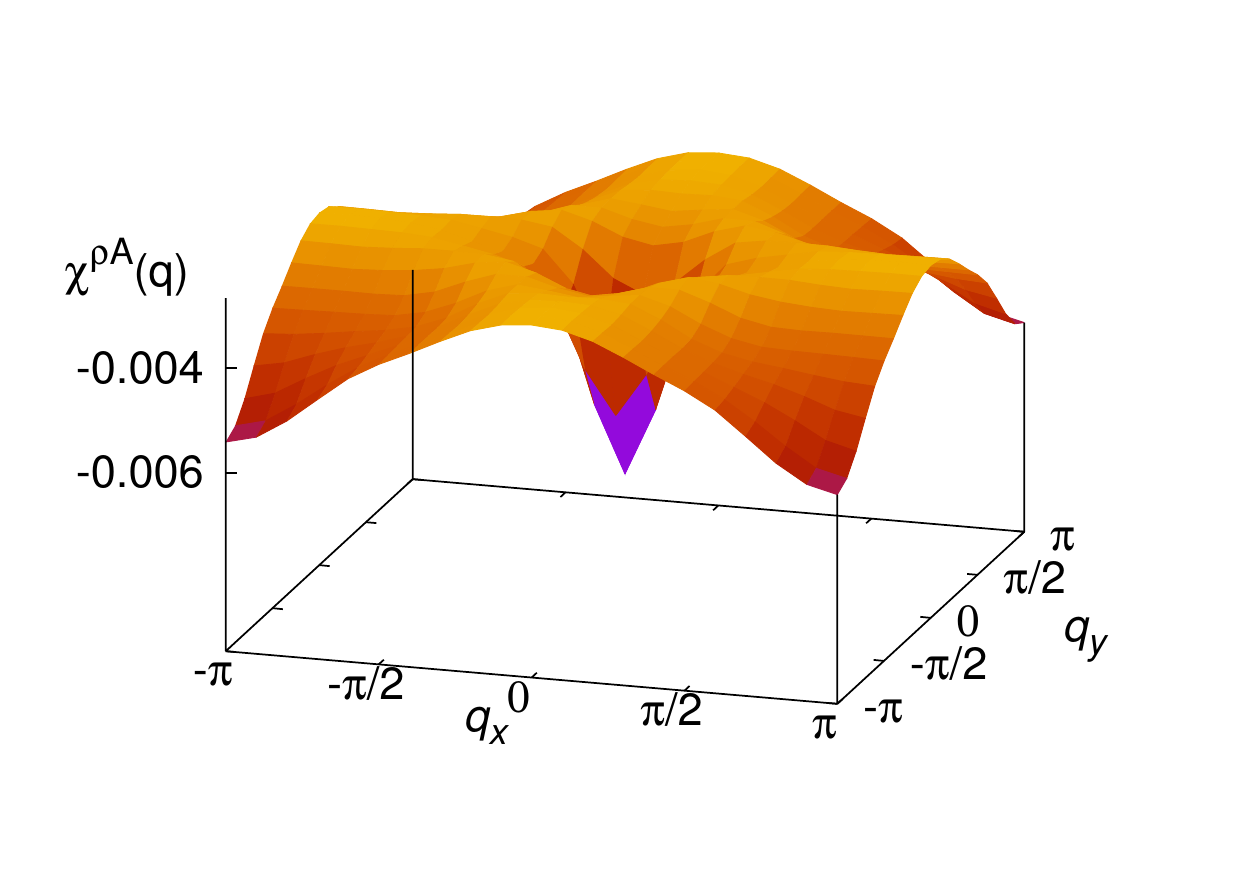}
\caption{(Color online) Average of Fourier transformed 
off-diagonal correlations $\chi^{A\rho}({\bf q})$ for
$V_0/t=2.0$ and $|U|/t=2$.}
\label{fig2a}
\end{figure}

Recently\cite{kaml13} the spatial dependence of the STM spectra in 
strongly disordered NbN films has been analysed in terms of the 
autocorrelation function for the order parameter, i.e.
\begin{equation}\label{eq:autocorr}
\langle C(\bR)\rangle=\frac{1}{N}\langle \sum_i
\left( \Delta_i-\langle\Delta\rangle \right)
\left( \Delta_{i+\bR}-\langle\Delta\rangle \right)\rangle.
\end{equation}
By performing an average over several disorder configurations we can extract
the corresponding correlation length $\lambda_{ac}$ from a fit to the
function
\begin{equation}\label{fit2d}
F(\bR)=a_0 + a_1 \mbox{e}^{-R/\lambda_{ac}\left( 1+a_2\sin^2(2\phi)
+a_3\sin^2(4\phi)\right)}\,.
\end{equation}
Here $\phi$ is the polar angle related to $\bR$ which incorporates
anisotropies in the correlations and we restrict the fit
to $|\bR|>2$ in order to isolate the long-distance behavior.
The resulting length $\lambda_{ac}$ as a function of disorder 
is shown by circles in the right inset 
to Fig.~\ref{fig4} and it is close to the correlation length $\xi_0^A$
extracted from the amplitude correlations. 
For $V_0\rightarrow 0$ one can apply linear response
  theory on the disorder and show that the two lengths coincide. In the strongly disordered regime the situation is more complex. We find numerically
  that both lengths are close to each other.
Notice that for $|U|/t=5$ we
observe that both $\lambda_{ac}$ and $\xi_0^A$ increase in the regime 
where the separation between the order parameter and the spectral gap 
starts to develop,   while they collapse in the regime where the 
spectral gap tends to increase again. In Monte Carlo 
simulations\cite{boua11}  the latter regime corresponds
  to the SIT, not captured by the present Bogoliubov-de-Gennes
  approach. This same tendency is observed in the experimental
  estimate of $\lambda_{ac}$ given in Ref. \onlinecite{kaml13}, done for samples
  in the so-called "pseudogap" region of the phase diagram, where the
  spectral gap is much larger than $T_c$.

\subsubsection{Momentum structure of off-diagonal correlations}
Off-diagonal correlations $\chi^{A\rho}({\bf q})$
mix the density and amplitude sector and are shown in
Fig.~\ref{fig1} for the clean case and in 
Fig.~\ref{fig2a} for the disordered system.

Upon coupling an external field in the density sector
$H_1=\sum_{\bf q} \lambda_{\bf q}\rho_{\bf -q}$ the correlation function $\chi^{A\rho}({\bf q})$
yields the corresponding response for the gap amplitude.
In particular, for ${\bf q}=0$ a spatially constant (and positive)
$\lambda_{{\bf q}=0}$ induces an effective reduction of the chemical potential.
Consider now the clean case where for the attractive Hubbard model
with nearest-neighbor hopping the gap amplitude as a function
of density has a maximum at half-filling and continuously
decreases towards $n=0$ and $n=2$. Therefore off-diagonal correlations
are negative for $n<1$ (where a positive $\lambda$ shifts the effective
chemical potential away from half-filling) in agreement
with Fig.~\ref{fig1} and positive for $n>1$. Similar arguments
can be made for finite momenta. In particular, the strong
enhancement of $|\chi^{A\rho}({\bf q})|$ at ${\bf q}={\bf Q}_{CDW}$
observed in Fig.~\ref{fig1} is due to the strong competition 
between CDW and SC correlations close to half-filling.

In the doped system Fig.~\ref{fig2a} reveals a strong suppression
for the off-diagonal correlations due to the spatial separation
of density- and amplitude fluctuations as demonstrated in 
Sec.~\ref{sec:decoup}. Naturally this is again most pronounced
for the CDW momentum due to the removal of particle-hole
symmetry by disorder. It is worth noting that in the dynamic limit (${\bf q}=0, \omega$ finite) 
the off-diagonal correlations show instead the opposite behavior. More specifically, as it has been recently discussed in Ref. [\onlinecite{cea_cm15}], the coupling between the amplitude and density/phase correlation at finite frequency is strongly enhanced by disorder, leading to a strong mixing between the amplitude and phase spectral functions at zero momentum.

\section{Current correlations}
To conclude our analysis of the SC correlations we shall discuss now the change in the current-current correlation function induced upon entering the
superconducting state. In particular we want to explore the
consequences of the percolative current formation (cf. Fig.\ \ref{fig2}) 
on the behaviour of the current
correlation function $\chi^{jj}$ entering the definition (\ref{eq:ja})
of the superfluid stiffness. 

In order to obtain the
intrinsic superconducting response, we have to subtract the contribution 
which is already present in the normal
state (at finite momenta), and which can be either diamagnetic or paramagnetic
depending on the filling of the system. 

\begin{figure}[htb]
\includegraphics[width=8cm,clip=true]{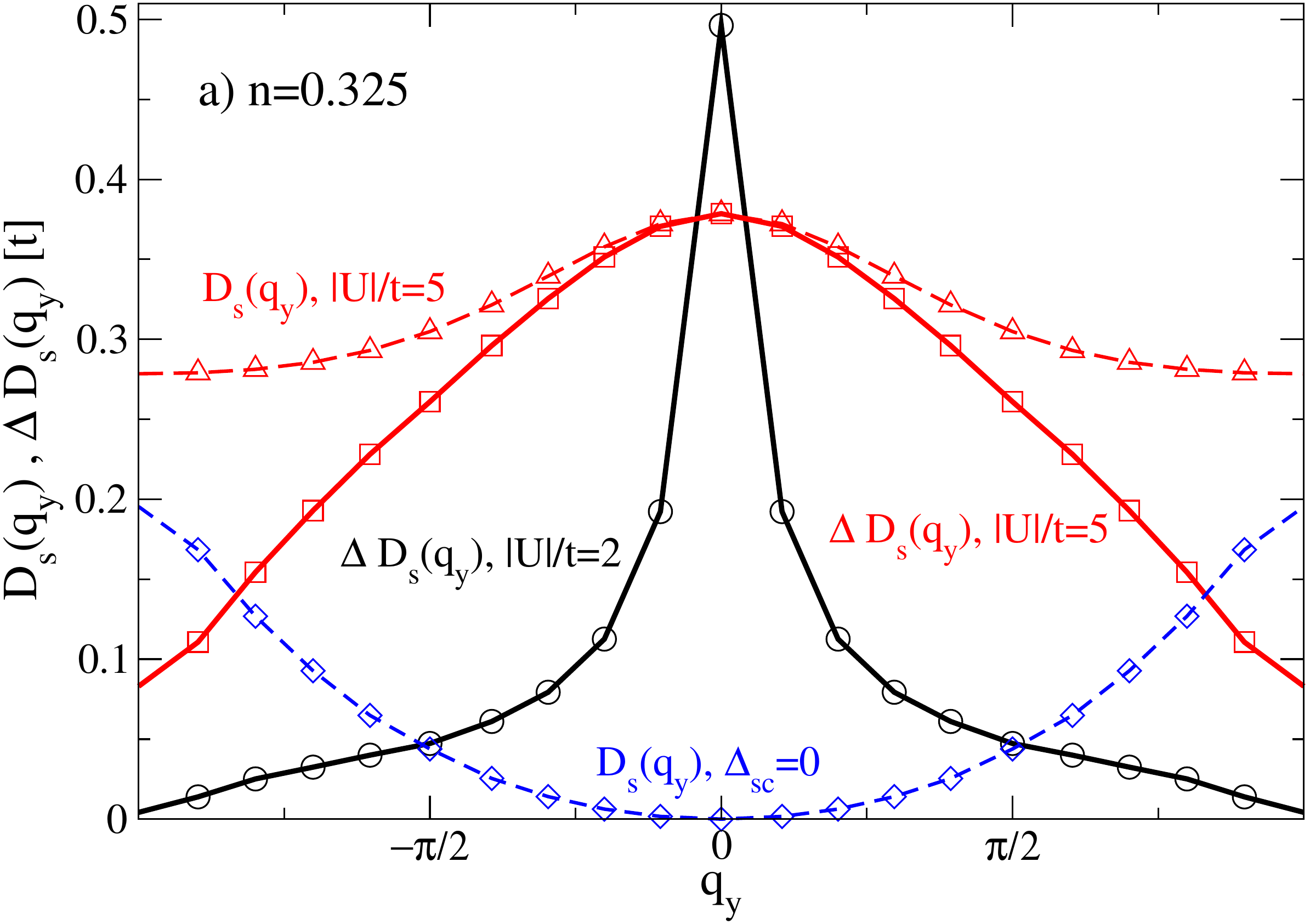}
\includegraphics[width=8cm,clip=true]{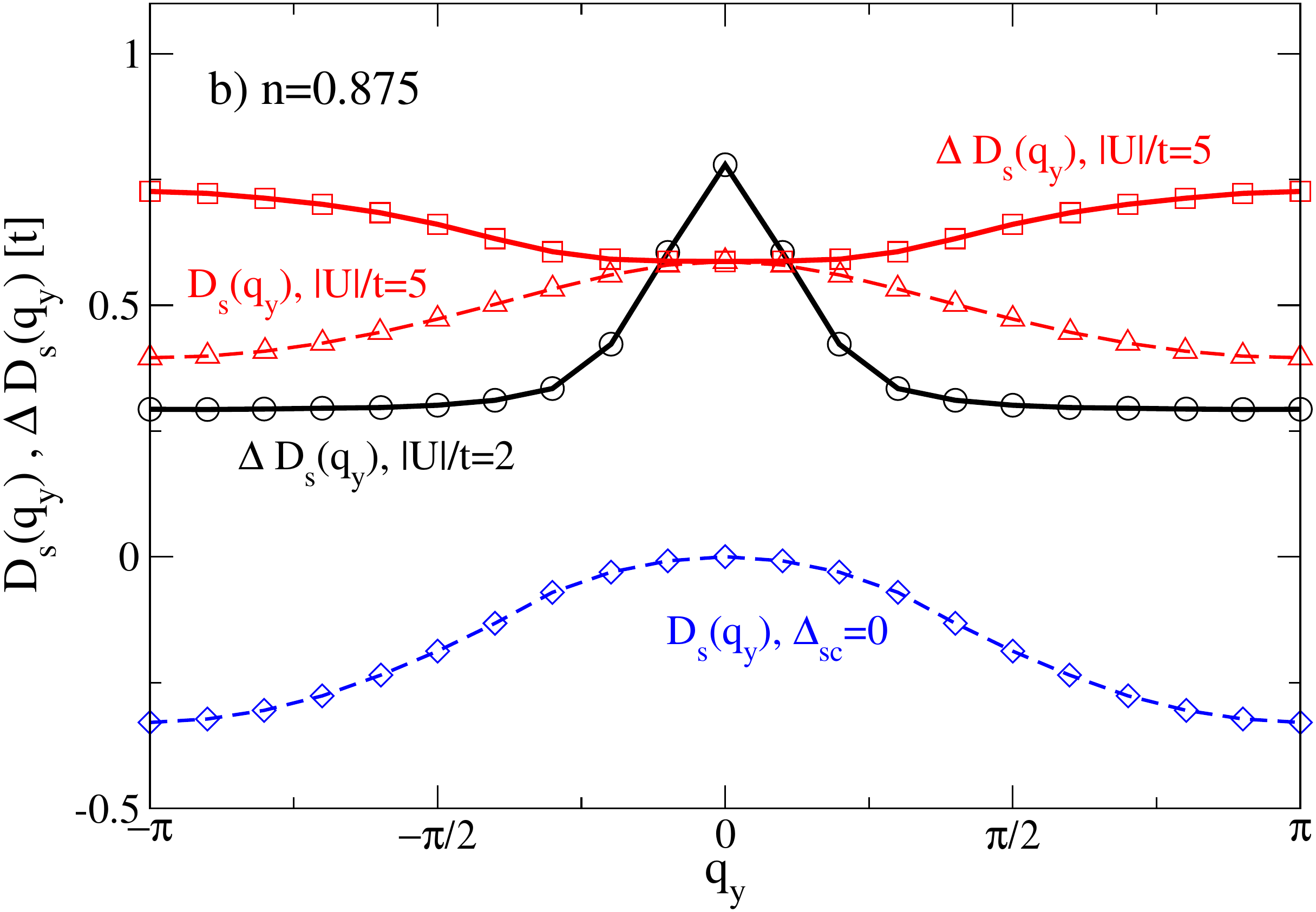}
\caption{(Color online) Transverse current response
$D_{q_y} = -\langle T_x\rangle + \langle \chi^{jj}({q_y})\rangle $ 
for the non-SC (blue dashed, diamonds) and the sc homogeneous 
system at $|U|/t=5$
(red dashed, triangles). The normal state response ($\Delta_{sc}=0$) is 
independent of $|U|/t$.
The solid lines report the difference
$\Delta D_s(q_y)$ between $D_{q_y}$ for the sc- and normal system 
for $|U|/t=5$ (squares)
and  $|U|/t=2$ (circles).  
Filling $n=0.325$ (a) and $n=0.875$ (b). }
\label{fig9a}
\end{figure}

This is illustrated by the dashed line marked with
diamonds in Fig.~\ref{fig9a} for the
homogeneous non-superconducting system.
Clearly, the current response of Eq.~\eqref{eq:jd}, 
$D_{q_y} = -\langle T_x\rangle + \langle \chi^{jj}({q_y})\rangle$
vanishes at $q_y=0$ (i.e. the SC stiffness) when the system is in 
the normal state, however, it becomes non-zero for finite momenta.

In particular at low density (cf. Fig.~\ref{fig9a}a) one recovers the 
finite-{\bf q} diamagnetic response ($D_{q_y}>0$) related to Landau 
diamagnetism 
in agreement with the transverse current response of a Fermi 
liquid.~\cite{pinesbook} In contrast, larger filling (cf. Fig.~\ref{fig9a}b)
supports a finite-{\bf q}
paramagnetic current response which would even diverge at ${\bf
  q}=(\pi,\pi)$  for $n=1$ (not shown).  
This feature is the starting point for the exploration of circulating
current phases as possible candidates for the pseudogap in 
cuprate superconductors.~\cite{schulz89}

As shown by the triangle symbols in Fig.~\ref{fig9a}
a finite SC gap shifts up the curves in order to yield a
diamagnetic $D_{q_y}$ independently on doping. In order to
extract what is due to superconductivity we take the difference
with respect to the normal state response $D^{normal}_{q_y}$ 
and the corresponding
curves are shown by square symbols ($|U|/t=5$) and circles  
($|U|/t=2$) in Fig.~\ref{fig9a} for $n=0.325$ and $n=0.875$, respectively.
In the weak coupling limit the difference 
$\Delta D_s(q_y)=D^{SC}_{q_y} - D^{normal}_{q_y}$ is always strongly
peaked at $q_y=0$ and the underlying normal state response does
not influence the curvature of the peak which determines the
SC coherence length. On the other hand, it turns out that for large
filling and strong coupling (cf. squares in Fig.~\ref{fig9a}) 
$\Delta D_s(q_y)$ can even acquire a maximum at the zone boundary.
Thus in this limit the SC diamagnetic response is largest on short
length scales and corresponds
to an oscillatory decay of the SC induced
current correlations in real space.

\begin{figure}[htb]
\includegraphics[width=8cm,clip=true]{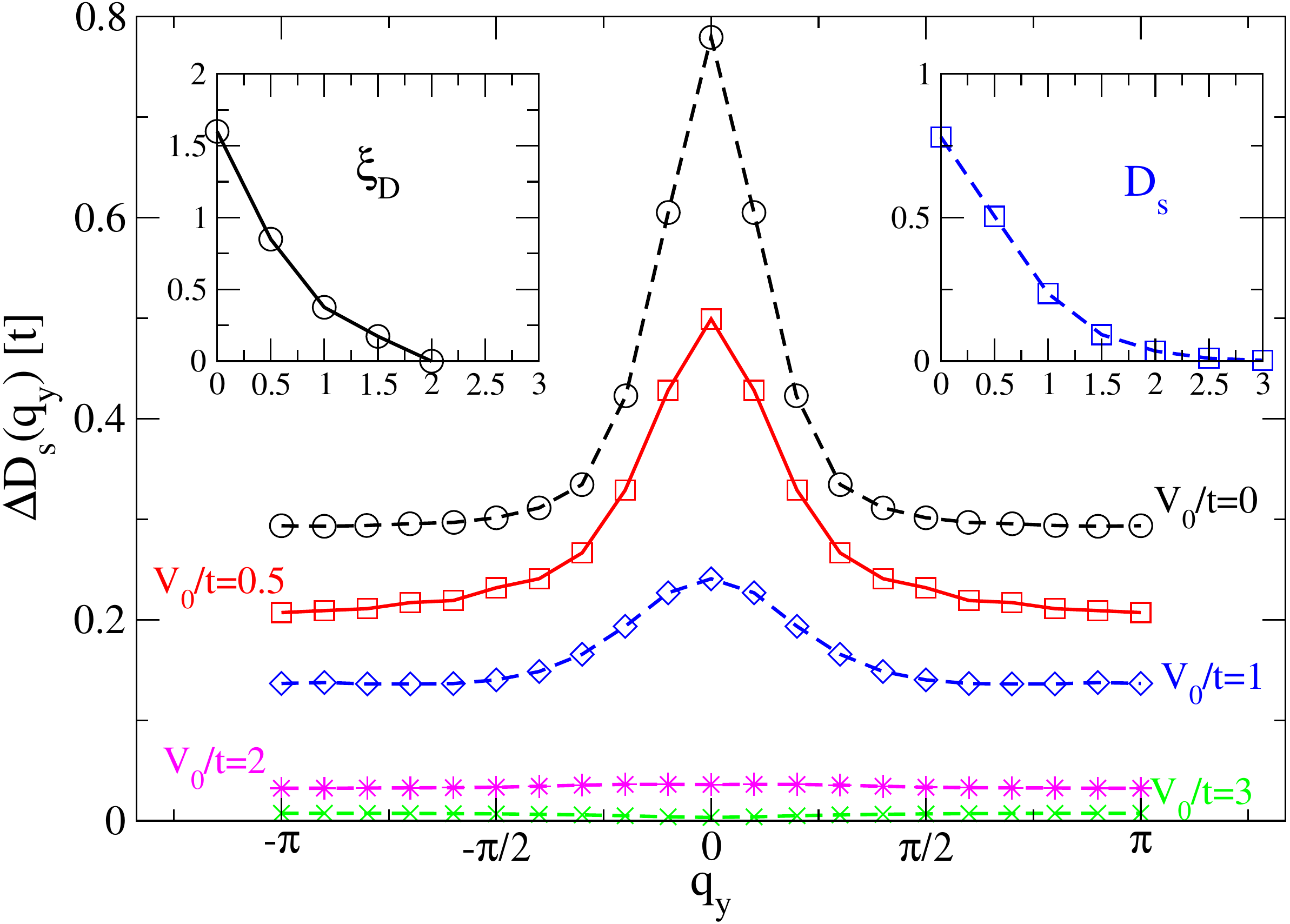}
\caption{(Color online) Main panel:  Transverse current correlations $\Delta D_s(q_y)$
measured with respect to the normal system for $|U|/t=2$, $n=0.875$, 
and various disorder strengths.
Upper left inset: Correlation length extracted from Eq. (\ref{eq:xi}).
Upper right inset: Superfluid stiffness. 
}
\label{fig9}
\end{figure}

Fig.~\ref{fig9} shows the transverse current response 
$\Delta D_s(q_y)$ for various 
disorder strength and interaction $|U|/t=2$ with the normal 
state result substracted. The latter has obtained for the
same disorder configurations and by setting $\Delta^{SC}_i=0$.

We parametrize the long-wavelength structure as
\begin{equation}\label{eq:xi}
\Delta D_s(q_y)=D_s\left[1-(\xi_D q_y)^2\right]
\end{equation}
which defines a SC coherence length related to the diamagnetic
response and allows us to extract the stiffness $D_s$ as a function of
disorder. Both quantities are shown in the insets to Fig.~\ref{fig9}.

As discussed previously \cite{sei12} (see also
Sec.~\ref{sec:mean-field-solution}) 
$D_s$ gets rapidly suppressed
with disorder but since the BdG approach does not capture the SC-insulator
transition it does not vanish even for large $V_0/t$. 
Also the coherence length (cf. left inset to Fig.~\ref{fig9})
is strongly suppressed by disorder. Above $V_0/t\approx 2$,  
$\Delta D(q_y)$ is essentially independent on the transverse
momentum $q_y$ and $\xi_D \approx 0$ within the numerical accuracy.

However, due to the average over disorder configurations the above 
analysis does not capture the long-range current
correlations which exist along the percolative path (cf. Sec.~\ref{sec:decoup})
and which we will analyze separately in the following.

\begin{figure}[htb]
\includegraphics[width=8cm,clip=true]{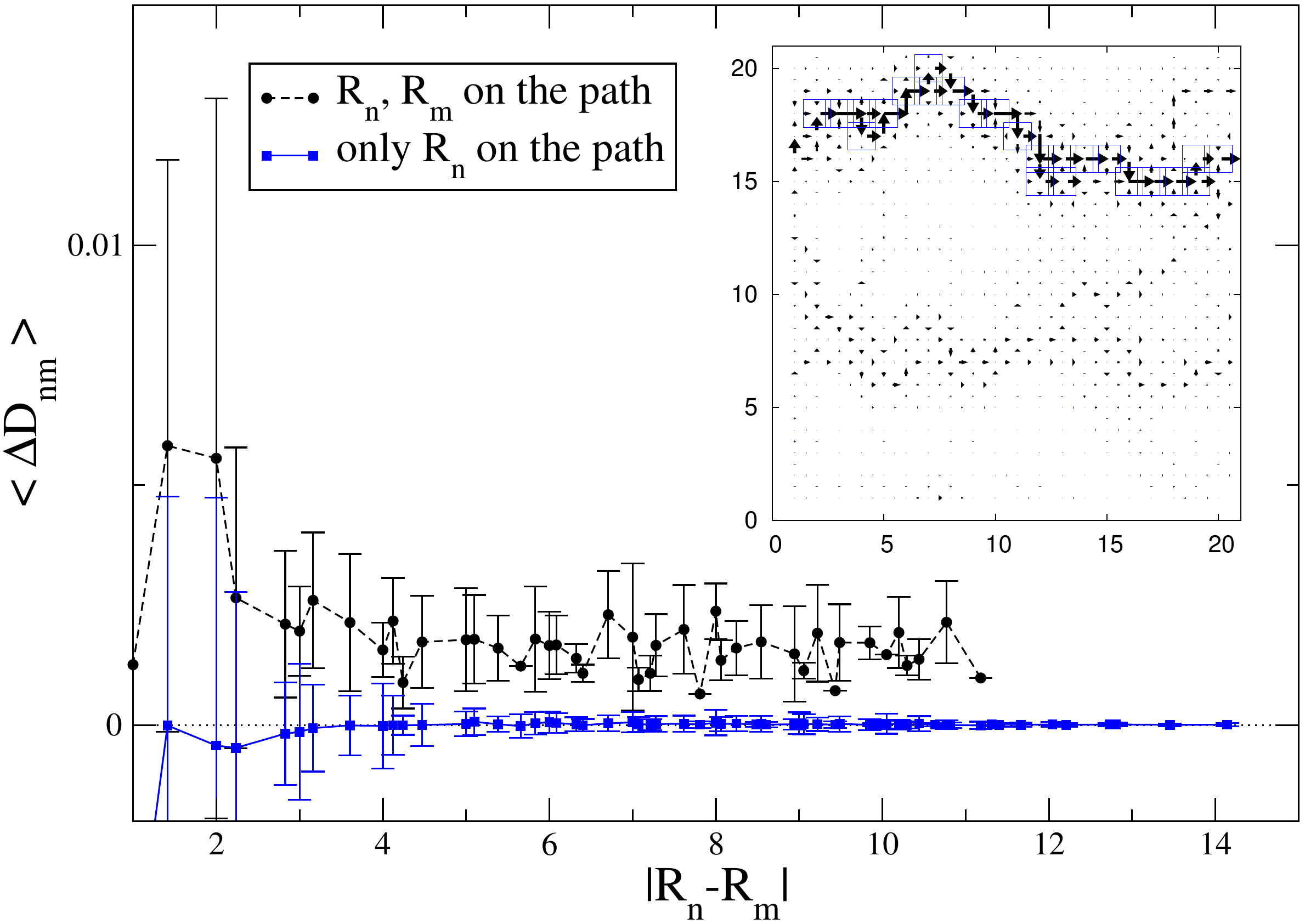}
\caption{(Color online) Main panel: $\langle \Delta D_{nm}\rangle $ for both sites
(black) and only one $R_n$ (blue) on the percolative path shown
in the inset. The squares indicate sites $R_n$ with $j_x(R_n) > j_x^{max}/3$. 
$|U|/t=2$, $n=0.875$, $V/t=3$.}
\label{fig11}
\end{figure}

First we identify the superconducting backbone. 
The criterion to decide which sites belong to the percolative path
is chosen as follows: For the vector potential ${\bf A}$ along the $x$ direction, we determine
the maximum current through a bond 
$j_x^{max}$ in the system and select all sites which have currents larger than 
$\alpha j_x^{max}$. We find that usually a value of $\alpha=1/3$ is 
appropriate in order to selecting the sites which are visited by the
path. An example is shown in the inset to Fig.~\ref{fig11} where
the squares indicate sites with $j_x(R_n) > j_x^{max}/3$. Clearly, there
are sites (e.g. in the upper right corner) which are traversed by
a minor current but are left out by the '$\alpha=1/3$' criterion. 
Reducing further the value of $\alpha$ would also include these sites,
however, we note that the following results do not depend sensitively on the
value of $\alpha$. The effect of a larger (smaller) $\alpha$
is to add sites with larger (smaller) current to the path which
concomitantly slightly increases (decreases) the long-distance correlations
which are calculated below.

We proceed by evaluating the non-local stiffness $D_{n,m}$ 
between sites $R_n$ and $R_m$
\begin{equation} \label{eq:Dsl}
D^{xx}_{nm} = \left\lbrack -\delta_{n,m} t_x(n) 
- \chi_{nm}(j^x_n,j^x_m)\right\rbrack
\end{equation}
and compute the difference between sc and normal state
$\Delta D_{nm} = D^{sc}_{nm}-D^{nl}_{nm}$.
Two cases are considered: (a) both sites $R_n$ and $R_m$
belong to the percolative path and (b) only one of the sites $R_n$, $R_m$ 
is on the path.
The result for $D_{nm}$ in both cases is shown in Fig.~\ref{fig11}
for the particular percolative path displayed in the inset. The 'errorbars'
indicate the variance due to the fact that different sites 
$R_n$ and $R_m$ have the same distance $|R_n-R_m|$ but different
values for $D_{nm}$.

As can be seen the current correlations rapidly decay away from
the percolative path and are practically 'zero' for $|R_n-R_m|>3$.
On the other hand correlations {\it on} the path stay finite
up to the largest distances available in the system.

\begin{figure}[htb]
\includegraphics[width=8cm,clip=true]{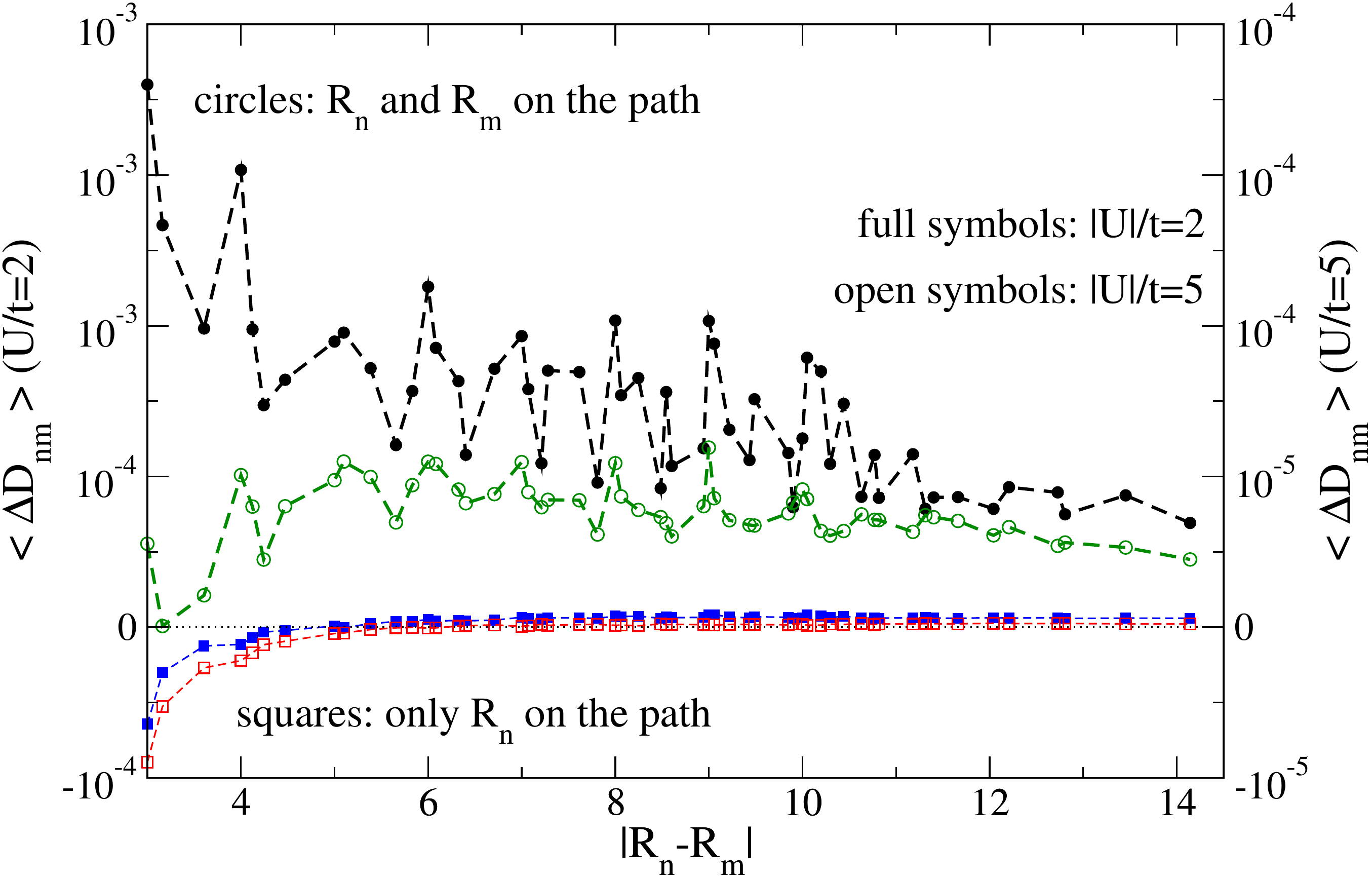}
\caption{(Color online) Current correlations 
$\langle \Delta D_{nm}\rangle $ averaged over 200 samples for
disorder strength $V/t=3$ and $n=0.875$. 
Full symbols: $|U|/t=2$; Open symbols: $|U|/t=5$.}
\label{fig12}
\end{figure}

Finally, Fig.~\ref{fig12} shows the on- and off-path current correlations
averaged over $200$ disorder configurations for $|U|/t=2$ and $|U|/t=5$, respectively. As for the specific
sample shown in Fig.~\ref{fig11} the off-path correlations get rapidly
suppressed while on-path correlations stay finite up to large $|R_n-R_m|$.
Upon comparing the on-path correlations between the two $|U|/t$-values
one finds, besides a reduction by a factor of $\approx 10$, 
that the decay of $\Delta D_{nm}$ with distance for $|U|/t=5$ is significantly
smaller than for  $|U|/t=2$ while still staying finite for the largest
possible separation in the system ($\approx \sqrt{2}\times 10$ for
a $20\times 20$ lattice when the percolative path is along the diagonal).

The persistance of the current correlations along the percolative path resembles closely the expected behavior for a one-dimensional chain, where it simply follows from the current conservation. This can be easily seen at $\omega=0$ by using the following classical phase-only action:
\begin{equation}\label{eq:gaussian}
S=\frac{1}{2} \sum_i  J_i (\delta \Phi_i)^2 
\end{equation}
where $J_i$ are the local (random) stiffnesses (in units of the temperature $T$) and $\delta\Phi_i$ represents the local phase gradient $\delta \Phi_i\equiv (\theta_{i+1}-\theta_i)$, $\theta_i$ being the local SC phase. Eq.\ (\ref{eq:gaussian}) can be obtained for example by expanding at Gaussian level a classical $XY$ model with random couplings $J_i$, that is the prototype model for the phase degrees of freedom of a superconductor. Eq.\ (\ref{eq:gaussian})  is also obtained\cite{cea14} by mapping\cite{ma_prb85} at large $U$ the disordered Hubbard model  into the pseudospin model. In this mapping  the superconductivity corresponds to
a spontaneous in-plane magnetization, i.e. to the usual $XY$ model with a coupling $J\sim t^2/U$, and disorder maps into a random out-of-plane field, that leads in turn to the disorder in the local couplings $J_i$ after a Holstein-Primakoff expansion around the 
mean-field solution.~\cite{cea14} 

The local current $I_i$ for the model (\ref{eq:gaussian}) can be written, after minimal coupling substitution $\delta \Phi_i\to \delta \Phi_i-2A_i$ in Eq.\ (\ref{eq:gaussian}) as:
\begin{equation}
\label{eq:i}
I_i=2J_i(\delta\Phi_i-2A_i)\,.
\end{equation}
In the one-dimensional case   the current conservation implies that $I_i$ is independent on the site index, i.e. $(\delta\Phi_i-2A_i)=c/2J_i$, where $c$ is a constant. By summing over the site index and using the boundary condition  $\sum_i \delta\Phi_i=0$ 
one then gets $c=-4(\sum_i A_i)/\sum_i(1/J_i)$. Since the superfluid stiffness is defines as usual (see Eq.\ (\ref{eq:jd})) as
$D_s=-I(q=0)/ A(q=0)$ one also deduces that 
\begin{equation}
D_{s}=4\left(\frac{1}{N}\sum_i\frac{1}{J_i}\right)^{-1}
\end{equation}
so that  $I_i= c=-(1/N)\sum_j D_s A_j$. By comparing this with Eq.\ (\ref{eq:Dsl}) above  we then recover that $D_{ij}=D_s/N$ for all pairs of sites $i,j$ along the chain. It is interesting to note that this result also implies that the paramagnetic contribution to the current must cancel out the local diamagnetic term $4J_i$ of Eq.\ (\ref{eq:i}).   This can be seen by computing explicitly the average current value from Eq.\ (\ref{eq:i}) in linear response theory, in analogy with the expression (\ref{eq:Dsl}) introduced above:
\begin{equation}
\label{eq:dsnl}
\langle I_i\rangle =-4\sum_jJ_i(\delta_{ij}-X_{ij}J_j)A_j\equiv -\sum_j D_{ij}A_j
\end{equation}
where $X_{ij}=\langle \delta\Phi_i\delta\Phi_j\rangle$ is easily determined from Eq.\ (\ref{eq:gaussian}) as:
\begin{eqnarray}
& &\langle \delta\Phi_i\delta\Phi_j\rangle=\nonumber\\
& &\frac{\int d\lambda
 {\cal D}\delta \Phi \exp\left[-\frac{1}{2} \sum_k  J_k (\delta \Phi_k)^2+i\lambda \sum_k \delta\Phi_k \right]
\delta \Phi_i\delta \Phi_j}{\int d\lambda
 {\cal D}\delta \Phi \exp\left[-\frac{1}{2} \sum_k  J_k (\delta \Phi_k)^2+i\lambda \sum_k \delta\Phi_k \right]}
 \nonumber\\
\end{eqnarray}
where the $\lambda$ integration accounts for the periodicity constraint. 
By making the change of variables $\delta\Phi_k\to \delta\Phi_k-i\lambda/J_k$ one immediately sees that 
\begin{equation}
X_{ij}=\frac{\delta_{ij}}{J_i}-\frac{D_s}{NJ_iJ_j}
\end{equation}
that inserted into Eq.\ (\ref{eq:dsnl}) gives $D_{ij}\equiv D_s/N$, 
as anticipated before. 
We note also that the independence of $D_{ij}$ in 
Eq.\ (\ref{eq:dsnl})  on both site indexes can be also derived as a 
consequence of charge conservation and gauge invariance in one dimension. 
Indeed the independence of $D_{ij}$ on the site index $i$ is a consequence of a constant current $I_i$ on each site,  while the independence of $D_{ij}$ on the second index $j$ is a consequence of the fact that at $\omega=0$ only 
to the $q=0$ component of the gauge field $A$ leads to a finite response.

Going back 
to our  2D system, we clearly see in Fig.~\ref{fig12}
that for a fixed disorder strength the percolative path 
becomes more '1D'-like with increasing $|U|/t$, 
which accounts for the crossover to a more constant $\Delta D_{nm}$
 for $|U|/t=5$. Indeed, a larger $|U|/t$ corresponds to a smaller $J$
in the mapping into the $XY$-like bosonic model, with an 
enhanced influence of disorder
and with smaller effective  local stiffnesses ${J_i}$.
This in turn is in agreement with the strong reduction of $\Delta D_{nm}$
from $|U|/t=2$ to $|U|/t=5$, as shown in Fig.~\ref{fig12}.

\section{Discussion and Conclusions}\label{sec:conc}

As we discussed in the introduction, it has been now established in several
theoretical models that when the SIT is approached a granular SC state
emerges, with SC puddles embedded in a non-SC background. Thanks to the
enormous progresses made in the experimental techniques able to probe the
systems in real space, it has been also established that such an emergent
granularity is observed in disordered films of 
conventional superconductors, like e.g. NbN, InO$_x$ and
TiN.~\cite{sac08,sac10,mondal11,chand12,kaml13,noat13} It is then crucial
to assess how this inhomogeneous SC state affects the behavior of the amplitude, density and current correlations, in order to interpret the results of the various experimental probes. 

\begin{figure}[htb]
\includegraphics[width=8cm,clip=true]{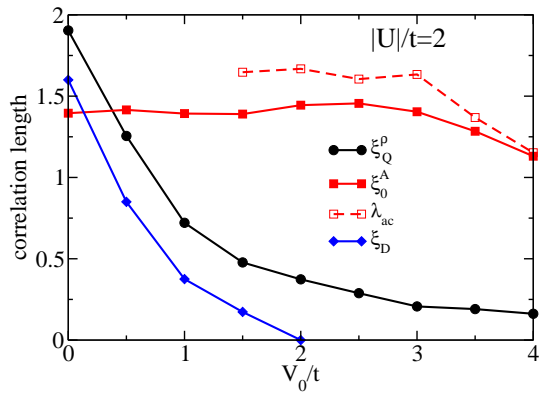}
\caption{(Color online) Summary of results for the various
  correlation lengths as a
  function of disorder for $|U|/t=2$ and $n=0.875$. At very small disorder the autocorrelation length $\lambda_{ac}$ cannot be properly defined since the approximated formula (\ref{fit2d}) does not reproduce accurately the data, see also $C(R)$ in Fig.\ \ref{figcr} below.} 
\label{figsum}
\end{figure}

In the present manuscript we analyzed this issue within the fermionic
Hubbard model with on-site disorder. We presented a detailed study of the
correlation functions  both in real space, for a specific disorder
configuration, and in momentum space, after the average over several disorder
configurations. The momentum-space analysis allows us to extract the
correlation length of each physical quantity in close analogy with the usual
approach for homogeneous systems. As a first result, one then sees that while
in the homogeneous case at low temperature amplitude and 
current correlation lengths coincide up to a numerical factor\cite{lara}, in
the presence of strong disorder this is no more true as can be seen in the
summarizing figure \ref{figsum}. By means of a simultaneous analysis of the
real-space correlations we can then disentangle how the properties of the
fragmented SC ground state
influence the various correlation lengths. As we discussed in the
manuscript, these two approaches give complementary informations, that
we will summarize below. In this respect, even though our results are
based on a RPA approximation, they have the advantage to allow  for
larger system sizes than Monte Carlo simulations, as e.g. those
reported in Ref.  \onlinecite{scal99}. The use of large clusters is in
turn crucial to trace back the behavior of different response
functions to the inhomogeneous structure of the ground state and to
perform a momentum-space analysis.

{\em Amplitude and density correlations}.

We find that in general the
strength of the amplitude response $\sim 1/(m_0^A)^2$ increases with
disorder while the charge response $\sim 1/(m_Q^\rho)^2$ gets
suppressed by disorder (cf. Figs.~\ref{fig6} and \ref{fig4}). This is
similar to a previous Monte Carlo study\cite{scal99} which found
that superconducting correlations are much more robust to disorder than
charge correlations. Here, due to the larger system size, we could explore
in detail the origin of this behavior.

The suppression of the charge response is
easily understood by the tendency of disorder to localize the pairs
and render the system incompressible almost everywhere except in the
superconducting islands. The increase of the superconducting response
is more subtle.  For strong disorder the region in between the islands
contains  ``marginal'' sites where the order parameter is small but
very susceptible to become large by small variations of the disorder 
[see Fig.~\ref{fig7}(b) and Fig.~\ref{fig8} for site (3,12)] yielding a
large overall pair susceptibility and  
resembling the behavior close to a second order phase transition. The
decoupling of density and amplitude correlations in real space is
reflected in the momentum-space structure of the
susceptibilities. Thus, while in the homogeneous case\cite{lara} the
maximum or $\chi_{\rho\rho}({\bf q})$ at the CDW vector
${\bf Q}=(\pi,\pi)$ leads to an enhancement of the amplitude correlations
$\chi_{AA}({\bf q})$  at the same wavevector (see Fig.\ \ref{fig1}),
in the disordered case this effect disappears (Fig.\ \ref{fig3}).

The resulting amplitude correlation length $\xi^A_0$, shown in
Figs. \ref{fig4}, \ref{figsum} has an interesting disorder dependence.
Indeed, it stays
constant or it is even enhanced   at intermediate disorder levels,
before then being ultimately suppressed as the SIT is approached. In
the latter regime we argued that the decay of the correlation length
is ruled by the behavior of the spectral gap, which
  increases as pairs become localized with disorder.

\begin{figure}[htb]
\includegraphics[width=8cm,clip=true]{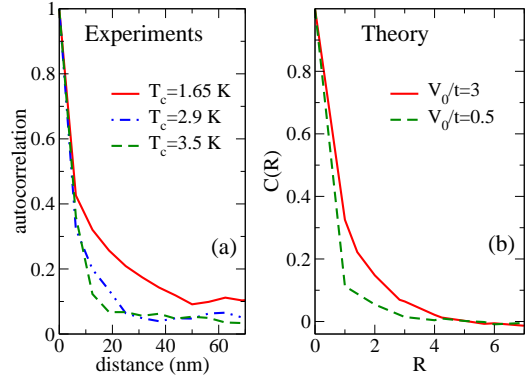}
\caption{Comparison between the experimental estimate  (left panel) of the autocorrelation function, defined in Eq. (\ref{eq:autocorr}), and the numerical computations (right panel). The experiments data are taken from Ref. [\onlinecite{kaml13}] and refer to three NbN films at different disorder level (labeled by the different critical temperatures $T_c$). The theoretical data are obtained for $|U|/t=2$, $n=0.875$, and 
disorder values $V_0/t=0.5$ (solid) and $V_0/t=3.$ (dashed). Considering that the typical size of the SC islands in these NbN films range between 20-40 nm, and it is one-two lattice spacings in our simulations, the length scales in the experiments and simulations are approximately comparable.} 
\label{figcr}
\end{figure}

In Figs.\ \ref{fig4}, we also compared $\xi_0^A$ with the
autocorrelation length $\lambda_{ac}$, that can
be directly extracted experimentally from the STM maps of the SC
ground state. This has recently been done for disordered NbN films \cite{kaml13}
and we show for convenience the corresponding data in Fig. \ref{figcr}a.
In this work the SC islands are identified by the regions with a
large SC coherence-peak height, that is usually taken\cite{boua11,lem13}
to be a measure of the local order parameter $\Delta_i$, i.e. the local gap
solution in the BdG equations. By analyzing the spatial correlations between
good SC sites the authors of Ref.\ [\onlinecite{kaml13}] found that the
autocorrelation length $\lambda_{ac}$ becomes {\em larger} as 
disorder is increased. This is shown in Fig. \ref{figcr}a where we report the experimental data for the autocorrelation function $C(R)$ defined in Eq.\  (\ref{eq:autocorr}) above. A similar trend can be observed also in our simulations, see Fig.\ \ref{figcr}b, where $C(R)$ shows first a rapid suppression over a length scale of the order of the SC island, followed by a long-tail decay that can be eventually fitted with the approximated formula (\ref{fit2d}) in order to extract $\lambda_{ac}$. Since this
tail can be thought as the response of the system to the fluctuations
that created the island we expect that  $\lambda_{ac}$ is close to 
$\xi_0^A$, as indeed we find numerically, see  Fig.\ \ref{fig4} and Fig. \ref{figsum}.

In contrast to the autocorrelation length, a direct estimate of the
  amplitude correlation length $\xi_0^A$  from the experiments is not so
  straightforward. Indeed, while within a Ginzburg-Landau approach,
  where a single length scale exists, $\xi_0$ can be estimated from
  the upper critical field at $T=0$ as $H_{C2}=\Phi_0/(2\pi\xi_0^2)$,  
at strong disorder this connection is not obvious. In particular when the 
superfluid stiffness $D_s$ is the lowest energy scale in the problem one 
would expect that  $T_c\propto D_s$, so that also the upper critical field 
will scale with $D_s$, as suggested for example by a recent analysis of the 
microwave conductivity at finite magnetic field in disordered 
InO$_x$.~\cite{armitage_prl13} In this sense, even though at intermediate 
disorder the decrease of $H_{c2}$ measured experimentally \cite{pratap11} can 
be interpreted as an increase of $\xi_0$ due to the weakening of the SC 
order parameter, as the SIT is approached one should not attribute the 
vanishing of $H_{c2}\propto T_c$ to a divergence of $\xi_0^A$ discussed above. 

{\em Current correlations} The behavior of the current correlations is also
strongly influenced by the formation of a fragmented SC state. Indeed, as
already noticed before,\cite{sei12} the superfluid response is mainly determined
by a few percolative paths that connect the good SC regions. As a consequence,
the decay of the current correlations depends on the position of the initial
and final sites with respect to this SC 'backbone'. If both sites belong to a
percolative path the current correlations are long-ranged  (essentially
constant, see Fig.\ \ref{fig11}),   in agreement
with what one expects for a truly one-dimensional system, like e.g. the
one-dimensional $XY$-model. On the other hand, this long-range behavior is
easily missed when the  transverse current correlations are extracted from
the response  in momentum space after average over several disorder
configuration. Indeed, the current-current correlation length $\xi_D$ is
rapidly suppressed (cf. inset to Fig.~\ref{fig9} and Fig. \ref{figsum}a), in
analogy with the overall superfluid response. This behavior has to be
contrasted to the one of the amplitude correlation length $\xi_0^A$, that is
strongly suppressed only at the SIT. On the other hand, the persistence of
current correlation along the percolative paths suggests that the existence of
the SC backbone can be deduced in principle by the measurements of the
space-dependent current susceptibilities, without having to evaluate
explicitly the current pattern at finite applied field. The experimental study
of these issues is of course challenging, but it should be accomplishable
with four-point
atomic force microscopy when the electrode spacing reaches the nanometer
separation. Its observation would certainly contribute significantly to our
understanding of the basic mechanisms leading to the formation of the
inhomogeneous SC state as the SIT is approached in real systems.

\acknowledgments
This work has been supported  by Italian MIUR under projects FIRB-HybridNanoDev-RBFR1236VV, PRINRIDEIRON-2012X3YFZ2 and Premiali-2012 ABNANOTECH, and by the  Deutsche
Forschungsgemeinschaft under SE806/15-1.

\begin{figure}[ttt!]
\includegraphics[width=8cm,clip=true]{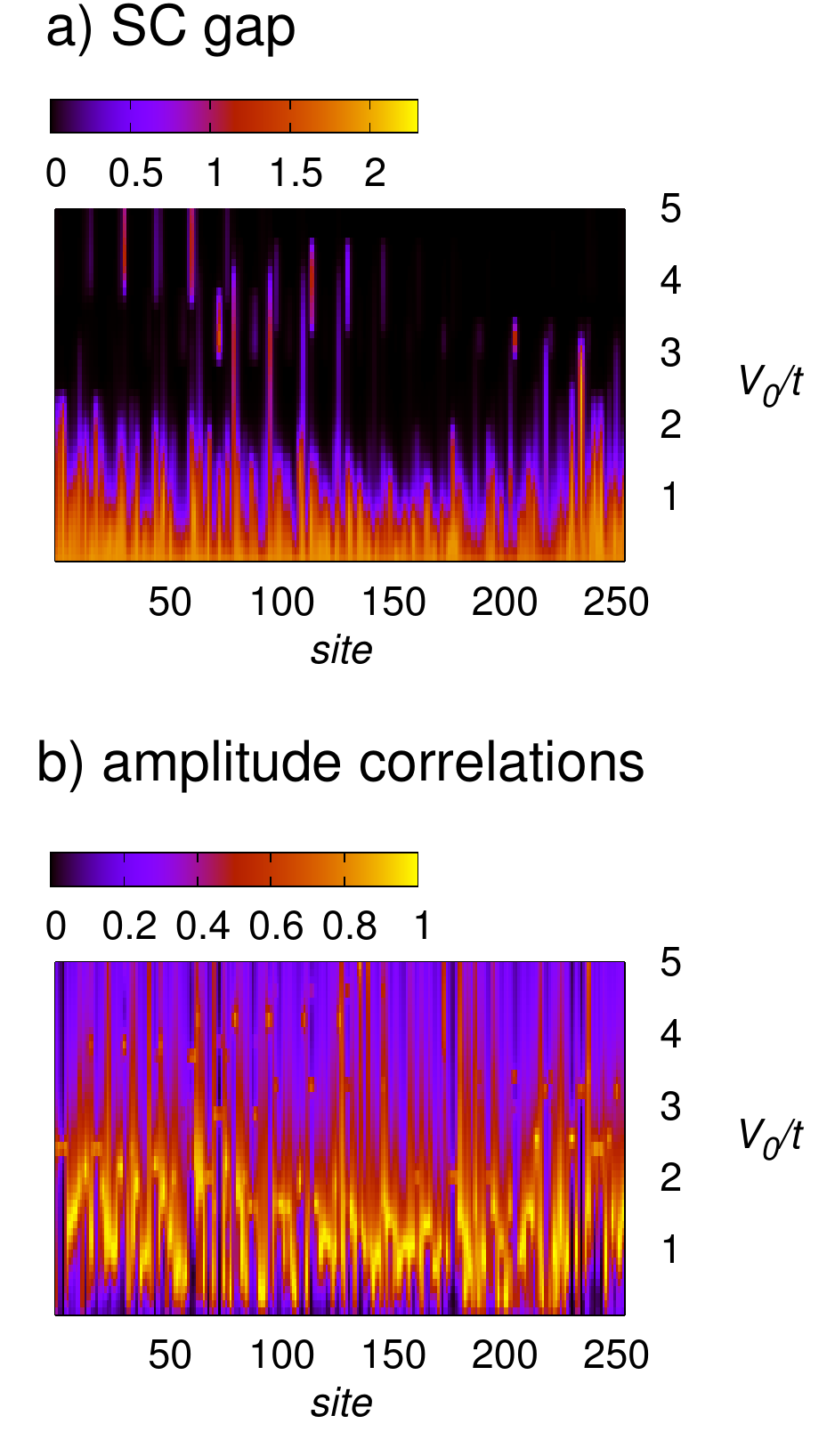}
\caption{(Color online) Top panel: disorder dependence of the SC gap value for
the same disorder configuration used in Figs. \ref{fig2}, \ref{fig7}.
The site index for the $16\times 16$ lattice is obtained
from $i_x+16(i_y-1)$.
Lower panel: disorder dependence of the local amplitude correlations 
$\chi^{AA}_{ii}$ normalized to their maximum value at each site.
$|U|/t=5$, $n=0.875$.}
\label{figapp1}
\end{figure}

\vspace*{0.5cm}

\appendix\label{apa}
\section{Disorder dependence of local SC gap and local correlations}
\begin{figure}[ttt!]
\includegraphics[width=8.5cm,clip=true]{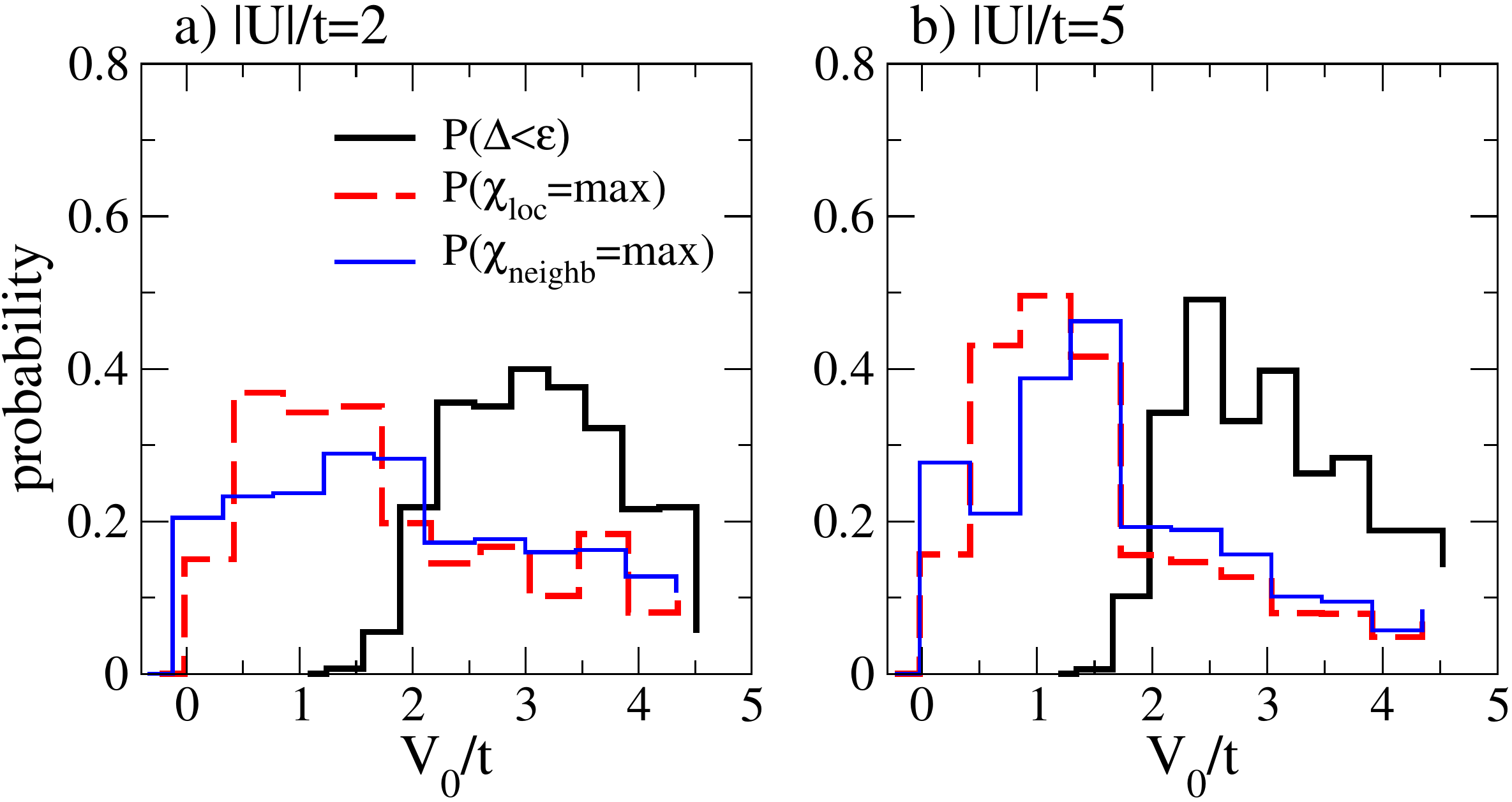}
\caption{(Color online) 
Black (full) steps: 
Probability distribution $P(\Delta<\epsilon)$ that the
  order parameter of a given site will
fall below the threshold $\epsilon$ for the first time, upon increasing the disorder with
$\epsilon=0.01t$ ; Red dashed step: 
Probability distribution $P(\chi^{AA}_{ii}=max)$ that the local amplitude
correlation of a given site will attain its maximum value as a function of
disorder strength.  Blue thin step: 
Probability distribution $P(\chi^{AA}_{\langle ij\rangle}=max)$ that the
nearest-neighbor amplitude correlations of a given bond  will attain its maximum value as a function of
disorder strength. 
Left panel: $|U|/t=2$, Right panel: $|U|/t=5$.}
\label{fig8b}
\end{figure}

Fig. \ref{fig8b} reports the disorder dependence of the 
SC gap value and local amplitude correlations on each site for
the same disorder configuration and parameters 
used in Figs. \ref{fig2}, \ref{fig7}. Note that the amplitude
correlations are normalized to their maximum value at each site.
Clearly, the SC order parameter on the majority of sites drops to
a small value around $V_0/t\approx 2$ but there are also
singular sites where $\Delta_i$ extends up to $V_0/t\approx 4$
or where $\Delta_i$ reemerges at large disorder values.

In the clean system the onset of a finite SC gap below $T_c$ is accompanied
by a divergence in the amplitude correlations, both the local and non-local 
ones.
The pronounced enhancement of the amplitude correlations around
$V_0/t \approx 2$ in Fig. \ref{figapp1}b suggests a similar feature as a 
function of disorder with the difference that $\Delta_i$ does not vanish but
becomes small beyond some value of $V_0$. To analyze this feature in more
detail we plot in Fig.~\ref{fig8b} the probability density
  $P(\Delta<\epsilon)$ as a function of the disorder strength $V_0$. 
Here $P(\Delta<\epsilon)dV_0$ is the probability that the order
parameter of a given site will fall below the threshold $\epsilon$ for
the first time when the disorder is increased from $V_0$ to  $V_0+dV_0$. 
Also shown are the probability distributions for the maximum
in the local [$P(\chi^{AA}_{ii}=max)$] and nearest neighbor 
[$P(\chi^{AA}_{\langle ij\rangle}=max)$] amplitude correlations where, 
for example, 
$P(\chi^{AA}_{ii}=max) dV_0$ is the probability that $\chi^{AA}_{ii}$ 
for a given site $i$, attains its maximum value as a function of disorder in the
interval $V_0$, $V_0+dV_0$. Clearly, for $|U|/t=5$ (right panel of
Fig.~\ref{fig8b}) $P(\Delta<0.01t)$ has a pronounced peak around 
$V_0/t\approx 2 \dots 2.5 $
and one finds that for about $50\%$ of all sites $\Delta_i<0.01t$
between $1.5 < V_0/t < 2.5$. Concomitantly also the probability
distributions for the local and non-local amplitude correlations
are peaked at a somewhat lower value of $V_0/t \approx 1.5$.
For smaller $|U|/t=2$ these distributions are broader and
in particular the nearest-neighbor amplitude correlations are
no longer characterized by a significant enhancement.

This finding offers an alternative perspective for understanding
the disorder dependence of the amplitude correlation length
$\xi_0$ shown in Fig. \ref{fig4}. Since $\xi_0$ is of the order of 
one lattice spacing
the nearest-neighbor correlations yield the dominant contribution 
to the correlation
length which accounts for the  enhancement around $V_0/t= 2$.
On the other hand, the distributions as a function
of $V_0/t$ are significantly broader
for $|U|/t=2$ (cf. Fig. \ref{fig8b}a) which agrees with the
behavior of $\xi_0$ shown in Fig. \ref{fig4}a.


\begin{thebibliography}{cc}
\bibitem{note1} Note that our definition for the susceptibilities differs
from those in Refs.~\onlinecite{strinati96,lara} by a factor $U/\chi_q^{0}$.
\end{thebibliography}
\end{document}